\documentclass[letterpaper,12pt]{article}
\usepackage[left=2cm,right=2cm,top=2cm,bottom=2cm]{geometry}
\usepackage{tensor}
\usepackage{calligra}
\usepackage{physics}
\usepackage[]{hyperref}
\usepackage{amsmath}
\usepackage{amsfonts}
\usepackage{pdfpages}
\usepackage{amssymb}
\usepackage{color}
\usepackage{mathrsfs}
\usepackage{latexsym}
\usepackage{physics}
\usepackage{cite}
\usepackage{graphicx}
\usepackage{float}
\usepackage{leftidx}
\usepackage{xcolor}
\usepackage{subfigure}
\usepackage{cleveref}
\usepackage{mathrsfs} 
\usepackage{caption}
\usepackage{tikz}
\usetikzlibrary{decorations.pathmorphing}
\graphicspath{ {images/} }
\title{Thin shell collapse in Kiselev geometry}
\author{ R. Saadati$^1$, F. Shojai$^{1,2}$\\$^1$Department of Physics, University of Tehran,\\Tehran, Iran.\\$^2$Foundations of Physics Group, School of Physics,\\Institute for Research in Fundamental Sciences (IPM),\\Tehran, Iran.\\}
\date{}

\begin{document}

\maketitle

\begin{abstract}
We present some new aspects of Kiselev black hole and then study the null and  timelike thin shell collapse in this space--time. For the latter, we show that Kiselev black hole can be matched to de Sitter core with a thin timelike  dust shell to produce a non-singular space--time. It is argued that for timelike hypersurface, the equation of state parameter must be non-negative. Using Barrabes-Israel junction conditions, the equation of motion of the shell is obtained. The stability of stationary solutions of the shell is discussed and some appropriate ranges for the parameters of shell and Kiselev geometry are found for which a stable stationary black hole is constructed.  
\end{abstract}
\section{Introduction}
Existence of black holes (BHs), as one of the predictions of general relativity, have drawn many attentions in theoretical physics\cite{Schwarzschild:1916uq}. Observing the first image of a BH in the nearby radio galaxy, M87, by the event horizon telescope collaboration \cite{Akiyama:2019cqa} for the first time, makes the subject even more interesting. Although it is almost clear to empirical physicists that BHs exist, the interesting problem of their inner structure is not well known. This is partly because of the masking effect of the event horizon. It is believed that the non-eternal BHs can be formed as a consequence of gravitational collapse of a star. According to the Hawking-Penrose singularity theorem \cite{Penrose:1964wq}, in general relativity, the gravitational collapse of reasonable matter leads to geodesically incomplete (i.e. singular) space--time such that this singularity remains hidden behind the event horizon. This is the result of weak cosmic censorship conjecture. Therefore, imposing some exotic conditions on matter \cite{Zaslavskii:2010qz} or concerning other extended theories of gravity \cite{Olmo:2012nx}, the BH singularity may be avoided. 

Inspired by Sakharov's work, who proposed the idea of replacing the Schwarzschild singularity with de Sitter vacuum \cite{Sakharov:1996}, Bardeen introduced the first ever regular BH \cite{Bardeen:1968}. 
Bardeen solution describes a static spherically symmetric space--time where for small (large) enough radial coordinate, approaches de Sitter (Schwarzschild) space--time.  Coupling Einstien equations to a new nonlinear electrodynamics, Ayon-Beato and Garcia \cite{AyonBeato:1998ub} generated Bardeen BH from a nonlinear magnetic monopole \cite{AyonBeato:2000zs}. Also, they proposed \cite{AyonBeato:1998ub} a non–singular exact BH solution where its corresponding source is a nonlinear electrodynamics satisfying the weak energy condition and in the weak field limit becomes the Maxwell field.  Later on, Bronnikov \cite{Bronnikov:2000yz} demonstrated that general relativity coupled to some nonlinear electrodynamics where the Lagrangian is a well-defined function of the Maxwell lagrangian, leads to a regular metric if and only if the electric charge is zero. This is in the case that the lagrangian has a correct weak field limit and tends to a finite limit as Maxwell lagrangian goes to infinity \cite{CQG}. This means that regular solutions can exist with a non zero magnetic charge. An ineresting minimal model of BHs of this type\footnote{For a classification of different types of regular, asymptotically flat, static and spherically symmetric BHs see \cite{Bronnikov:2006fu}.} is Hayward BH \cite{Hayward:2005gi}. Other similar proposals of regular BHs are found in \cite{Dymnikova:1992ux}. 

The above mentioned regular BHs models are described by regular solutions where the metric smoothly tends to the de Sitter one as $r\rightarrow 0$. These present globally regular space--times in which no junction conditions are needed in principle. However there are regular BHs constructed by joining two regions of space--time, the inner is described by a regular metric and the outer is a known BH solution. These are matched to each other by a smooth junction, boundary surface \cite{Mars: 1996,Magli:1997mw,Elizalde:2002yz,Conboy:2005nx}, or through a surface layer, thin shell \cite{Frolov:1989pf,Balbinot:1990zz,Lake:1989,Uchikata:2012zs,Masa:2018elb} which is of interest here. Using  Barrabes-Israel junction conditions \cite{Barrabes:1991ng}, two distinct space--times can be attached to each other with a timelike, spacelike or null hypersurface. Assuming some universal upper limit for the curvature of space--time, Frolov and collaborators \cite{Frolov:1989pf} proposed  a non-singular BH model by matching the Schwarzschild metric to a de Sitter one with a thin spacelike shell. They assumed that as the curvature reaches its upper value, the matter turns into a de Sitter phase and this transition is made through a spacelike thin shell. The stability of their solution is discussed by Balbinot and Poisson \cite{Balbinot:1990zz}.
Fitting of de Sitter space--time into a Schwarzschild BH with a spacelike surface layer of constant curvature is done in\cite{Lake:1989}. The intrinsic structure of the layer is obtained and it is shown that the fitting procedure can not be occurred through a boundary surface. As an important result, Poisson and Israel \cite{Poisson:1988wc} demonstrated that, due to the violation of junction conditions, Schwarzschild space--time cannot be matched directly to the de Sitter one with a null hypersurface and a thin shell is required. The matching is done later by Barrabes and Israel \cite{Barrabes:1991ng} and then discussed by many authors, see  \cite{Lemos:2011dq} for a detailed analysis. 

Another example is provided by matching a Reissner-Nordstrom BH to a regular de Sitter core \cite{Uchikata:2012zs} by a dust timelike thin shell such that at a specific radius, the transition between two space--times occurs. Then the stability of solutions is examined and it is shown that solutions with negative shell mass cannot be stable. Taking the massless limit of the shell, the result is the same as obtained before in \cite{Lemos:2011dq} with a boundary surface. Recently, this work is extended by considering a material layer with pressure in \cite{Masa:2018elb}.

In this paper, we employ Barrabes-Israel junction conditions to construct a new regular BH space--time. The outer metric is given by Kiselev BH \cite{Kiselev:2002dx} and the core is de Sitter space--time. Also the thin shell is chosen to be a dust timelike hypersurface. The outline of this paper is as follows: In the next section, we describe step by step how one can derive a generalized Kiselev metric. This comes from the fact that we have not restricted ourselves to linear equation of state of matter. In section \ref{Horizon}, we obtain some new information about the number and location of the Kiselev BH's horizons.  Then in section \ref{Null thin shell collapse}, the gravitational collapse of a null shell is studied in Kiselev background. Section \ref{Timelike thin shell collapse} is devoted to the main problem of the paper, the gravitational collapse of a timelike thin shell in Kiselev space--time where the interior space--time is de Sitter. After a brief review on the Barrabes-Israel junction conditions in \ref{Junction Conditions}, section \ref{The motion of a collapsing timelike shell} is devoted to derive  the equation of motion of the thin shell.  Section \ref{secRBH} deals with shell stability and in  section \ref{sec: Concluding Remarks}, we review highlights of the paper.

Throughout this paper, the signature of the metric tensor is assumed to be $(-,+,+,+)$. Greek indices ($\alpha$, $\beta$, ...) are
used to label the four dimensional space--time described by the metric components  $g_{\mu\nu}$ and Latin indices (a, b, ...) are reserved for objects live on the hypersurface $\Sigma$ defined by the three dimensional induced metric  $h_{ab}$. The symbol $;$ and $|$ are used to indicate  the covariant derivatives in four and three dimension respectively. A dot denotes the derivative with respect to the proper time. For any tensorial quantity like $A$ defined on both sides of $\Sigma$, the notation $[A]\equiv A|^{+}_{\Sigma}-A|^-_{\Sigma}$ assigns the jump of the $A$ across $\Sigma$. We use geometrized units where $c = G = 1$.
\section{Kiselev BH}\label{sec: Kiselev Space-time}
Kiselev metric firstly proposed in \cite{Kiselev:2002dx} to describe a static spherically space--time in the presence of an anisotropic
fluid except for the case of a cosmological constant where the mentioned fluid is isotropic.
It is a well-known fact that in the case of a spherically symmetric space--time of the form
\begin{equation}\label{eq: caconical form}
ds^2=-f(r)dt^2+\frac{1}{f(r)}dr^2+r^2d\Omega^2
\end{equation}
Einstein equations become linear in $f$ and give the non-vanishing components of energy-momentum tensor as
\begin{equation}\label{eq: EN eq1}
T\indices{_t^t}=T\indices{_r^r}=-\frac{1}{\kappa r^2}\left(f+rf'-1\right)
\end{equation}
\begin{equation}\label{eq: EN eq2}
T\indices{_\theta^\theta}=T\indices{_\phi^\phi}=-\frac{1}{2\kappa r}\left(2f'+rf''\right)
\end{equation}
The first equalities in (\ref{eq: EN eq1}) and (\ref{eq: EN eq2}), do not hold for a perfect fluid except for
the case of cosmological constant. To satisfy these equations, Kiselev's idea
is to construct an energy-momentum tensor via the following steps
\begin{itemize}
\item{Write a general spherically symmetric energy-momentum tensor in Cartesian coordinate system
\begin{equation} \label{eq: EM tensor in Cartesian}
T\indices{_\mu^\nu}=\left(\begin{array}{cc}
A(r) & 0 \\
\\
0    & C(r)r_i r^j+B(r)\delta\indices{_i^j}\\
\end{array}\right).
\end{equation} }
\item{Take its angular average 
\begin{equation}\label{eq: averaged EM tensor}
\langle T\indices{_\mu^\nu}\rangle=\left(\begin{array}{cc}
A(r) & 0 \\
\\
0    & \left(\frac{1}{3}r^2C(r)+B(r)\right)\delta\indices{_i^j}\\
\end{array}\right)
\end{equation}
and identify it with the energy-momentum tensor of a perfect fluid with energy density $\rho(r)$ and pressure $p(r)$
\begin{equation} \label{eq: ABCa}
A(r)=-\rho(r)
\end{equation}
\begin{equation}\label{eq: ABCb}
\frac{1}{3}r^2C(r)+B(r)=p(r).
\end{equation}}
\item{Write (\ref{eq: EM tensor in Cartesian}) in spherical coordinates by a coordinate transformation. The result will satisfy first equalities of (\ref{eq: EN eq1})
and (\ref{eq: EN eq2}) if 
\begin{equation} \label{eq: ABCd}
C(r)r^2+B(r)=-\rho(r).
\end{equation} }
\item{Read the unknown functions $A(r)$, $B(r)$ and $C(r)$ from (\ref{eq: ABCa})-(\ref{eq: ABCd}) and then find the energy-momentum tensor in spherical coordinates. This yields
\begin{equation}\label{eq: EM final}
{T}\indices{_{\mu}^{\nu}}=\text{diag}\bigg [ -\rho(r), -\rho(r), \frac{1}{2}(\rho(r)+3p(r)), \frac{1}{2}(\rho(r)+3p(r))\bigg ].
\end{equation} }
\item{Substitute $T_{\mu\nu}$ in (\ref{eq: EN eq1}) and (\ref{eq: EN eq2}), one can find the functions $f(r)$, $\rho(r)$ and $p(r)$. To do this, either one of these functions or a relation between two of them is needed. Specifying the equation of state,
$p=p(\rho)$ is an example of the latter case. In \cite{Saadati:2019cym}, the authors have used the equation of state of modified Chaplygin gas and found some analytical expressions for energy density, pressure and metric coefficient.}
\end{itemize}
Following the above mentioned steps, for a linear equation of state\\ $p(r)=\omega\rho(r)$, one finally arrives at the following expressions for the metric coefficient of Kiselev metric{ and the corresponding radial and transverse pressures 
\begin{equation}\label{eq: f_q}
f=1-\frac{2m}{r}-\frac{c}{r^{3\omega+1}}
\end{equation}
\begin{equation}\label{eq: Kiselev rho p}
\rho=-p_r = -\frac{3c\omega}{\kappa r^{3(\omega+1)}},\quad\quad p_t=-\frac{3c\omega(1+3\omega)}{2\kappa r^{3(\omega+1)}}.
\end{equation}
where $2m$ (i.e. the Schwarzchild radius), and $c$ are integration constants. Interestingly, one can find out that although the source fluid of Kiselev BH is anisotropic, i.e. $p_r\neq p_t$, the average pressure $\bar{p}$ satisfies a linear equation of state \cite{Visser:2019brz,Boonserm:2019phw}
\begin{equation}
\bar{p}=\frac{p_r+2p_t}{3}=\omega\rho.
\end{equation}
Note that for $\omega=1/3$ , the average pressure satisfies the radiation equation of state and the metric (\ref{eq: f_q}) reduces to the Reissner-Nordstrom solution.
The weak energy condition implies $\rho\geq0$, therefore from \eqref{eq: Kiselev rho p} the multiplication of $\omega$ and $c$ must be negative.
This also leads to a negative radial pressure in contrast to the transverse pressures.
Other feature of
Kiselev space--time is that the null energy condition is violated for $c\omega\left(\omega+1\right)>0$ \cite{Boonserm:2019phw},  hence the surrounding matter is exotic. 


In summary, the mentioned steps allow one to find a family of solutions each of which associated to a specific equation of state parameter of matter. Combining (\ref{eq: EN eq1}), (\ref{eq: EN eq2}) and (\ref{eq: EM final}), one gets
\begin{equation}
f=1-\frac{a}{r}-\frac{\kappa}{r}\int_{a}^r\! dr\ r^2\rho(r) ,\quad\quad \frac{r}{3}\frac{d\rho}{dr}=\rho+p
\end{equation}
where the horizon is assumed to be $r=a$. Above equations are linear so their solutions corresponding to the different energy densiy, can be superposed \cite{Pad}. For the case of a linear equation of state,  this means that for a sum of different sources with different values of state parameters, the corresponding coefficient of metric would be 
\begin{equation}\label{eq: Kiselev Metric}
f=1-\frac{2m}{r}-\sum_{n}\frac{c_n}{r^{3\omega_n+1}}
\end{equation}

\section{Horizons in Kiselev BH}\label{Horizon}
Here, we present a new qualitative description of the number and location of Kiselev BH's horizons. It is clear from  \eqref{eq: f_q} that one can not determine them for arbitrary values of $\omega$ analytically. Therefore many authors have addressed this issue by selecting some particular values of $\omega$. For the case of $\omega=-2/3$, a detailed analysis of null geodesics is done in \cite{Fernando:2012ue} and the structure of horizon is discussed in \cite{Fernando,Azreg-Ainou:2014lua,Ghaderi}. Moreover it is shown that this choice of $\omega$ gives a Nariai type BH  \cite{FERNANDO:2013uxa}. 

Introducing some dimensionless variables $u\equiv \frac{r}{2m}$ and $\tilde{c}\equiv \frac{c}{(2m)^{3\omega+1}}$, we note that the sign of $\tilde{c}$ is the same as $c$, $u$ is  positive and the metric component can be written as
\begin{equation}\label{eq: Horizon1}
f=1- \frac{1}{u}-\frac{\tilde{c}}{u^{3\omega+1}}.
\end{equation}
The horizon is now at $u=u_0=const$ where $\tilde{c}=u_0^{3\omega}(u_0-1)$. Combining this with the positiveness of the energy density condition  mentioned before, the multiplication of $\omega$  and $\tilde{c}$ must be negative. This gives
\begin{equation}\label{eq: Horizon5}
\omega(1-u_0)>0.
\end{equation}
Equation \eqref{eq: Horizon5} reveals that for $\omega>0$, the Kiselev BH's horizon(s) is (are) larger than $2m$ and vice versa. Moreover, note that the extremum of (\ref{eq: Horizon1}) is at $\tilde{u}=\left[-(1+3\omega)\tilde{c}\right]^{1/3\omega}$. Solving this for $\tilde{c}$ and substituting it into the condition $\tilde{c}\omega<0$, one  finds that
\begin{equation}\label{eq: Horizon3}
\frac{\omega}{3\omega+1}>0.
\end{equation}
Thus  $f$ has no extremum within $-1/3<\omega<0$ whereas for other values of $\omega$, it has exactly one extremum. Putting these all together, we can divide the parameter space,  $\tilde{c}$ and $\omega$, into different regions depending on the number of horizons and the positivity of energy density. First, consider the case that $-1/3<\omega<0$. For this interval, $\lim_{u\rightarrow 0^+}f(u)=-\infty$ and $\lim_{u\rightarrow\infty}f(u)=1$. Thereby, Kiselev BH has exactly one horizon. 
For the case of $\omega<-1/3$ or $\omega>0$, the BH has at most two horizons. The extremal case occurs once we have $f(u_0)=0$ and this means that 
$\tilde{c}$ has the following value \cite{FERNANDO:2013uxa}
\begin{equation}\label{eq: Horizon4}
\tilde{c}_{ext}=-\frac{1}{3\omega+1}\left(\frac{3\omega}{3\omega+1}\right)^{3\omega}.
\end{equation}
We have called the degenerate solutions for $w<-1/3$ "Nariai-Kiselev BH" because the BH solutions with two horizons 
 in this region of the parameter space are similar to that of the Schwarzschild-de Sitter solutions with the cosmological and event horizons, and thus the degenerate case is similar to the Nariai limit. The causal structure of this space--time is plotted in the conformal diagram (\ref{fig: Penrose Nariai}). It is evident that  any radial timelike observer falling from infinity in this space--time will either cross the horizon and reach the singularity or scape to one of the asymptotic points $\mathscr{P}$. For the corresponding conformal transformation and a rigorous discussion of the causal structure of extreme Schwarzschild-de Sitter space--time see  \cite{Podolsky:1999ts}.

If $\omega<0$ and $\tilde{c}>\tilde{c}_{ext}$, then $f(u)>0$ and there is a naked singularity. A similar argument can be applied when $\omega>0$ and $\tilde{c}<\tilde{c}_{ext}$.
Figure \ref{fig: Horizons} presents a summary of the results in the parameter space.
\begin{figure}[H]
	\centering
	\includegraphics[width=0.8\textwidth]{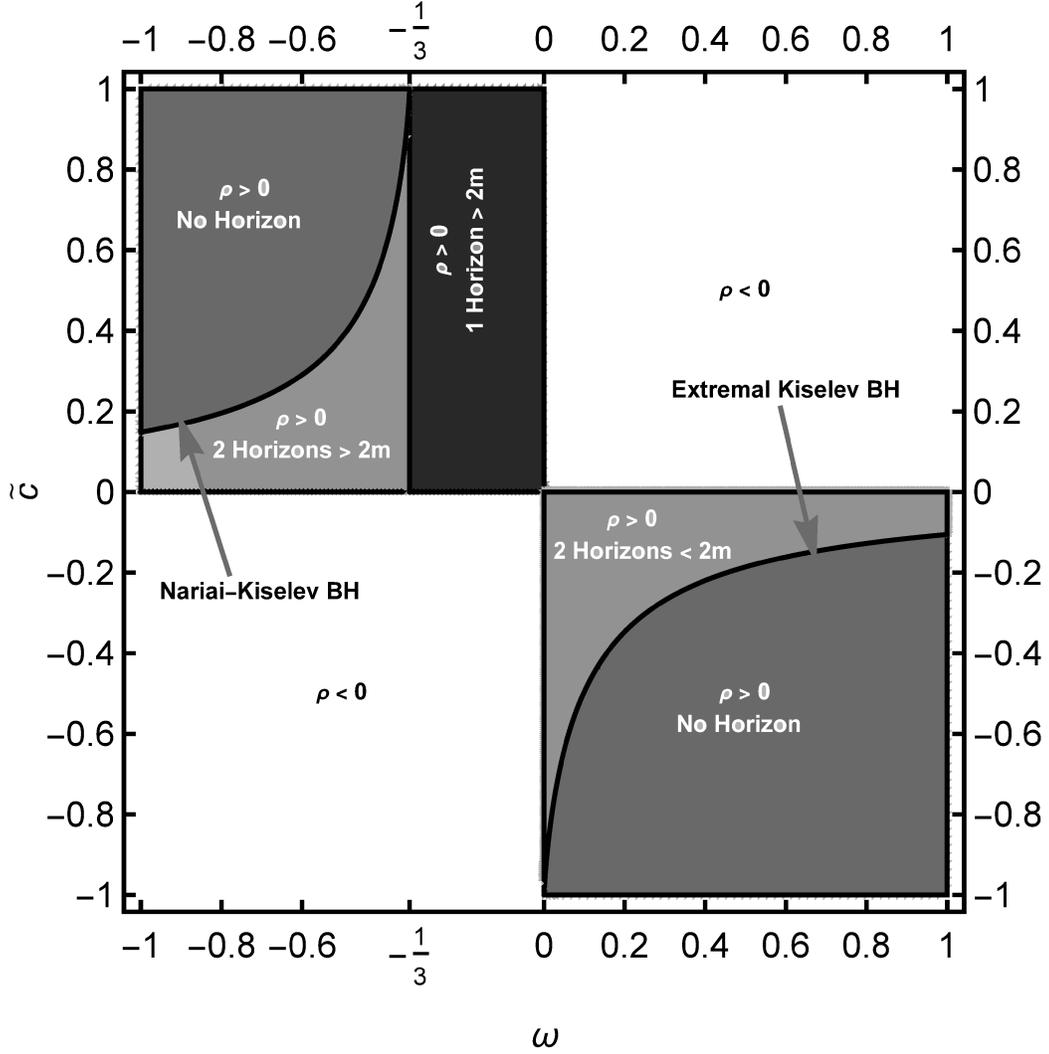}
	\caption{The properties of Kiselev space--time in parameter space ($\tilde{c}$, $\omega$). $\omega<-1/3$ correspond to Nariai-Kiselev BHs.}
	\label{fig: Horizons}
\end{figure}

\begin{figure}[H]
\centering

\tikzset{every picture/.style={line width=0.75pt}} 

\begin{tikzpicture}[x=0.75pt,y=0.75pt,yscale=-1.5,xscale=1.5]
	
	\draw    (101.66,248.67) -- (173.34,177) ;
	\draw [shift={(175,175.33)}, rotate = 315] [color={rgb, 255:red, 0; green, 0; blue, 0 }  ][line width=0.75]      (0, 0) circle [x radius= 3.35, y radius= 3.35]   ;
	\draw [shift={(100,250.33)}, rotate = 315] [color={rgb, 255:red, 0; green, 0; blue, 0 }  ][line width=0.75]      (0, 0) circle [x radius= 3.35, y radius= 3.35]   ;
	\draw    (176.67,176.98) -- (249.33,248.68) ;
	\draw [shift={(251,250.33)}, rotate = 44.62] [color={rgb, 255:red, 0; green, 0; blue, 0 }  ][line width=0.75]      (0, 0) circle [x radius= 3.35, y radius= 3.35]   ;
	\draw [shift={(175,175.33)}, rotate = 44.62] [color={rgb, 255:red, 0; green, 0; blue, 0 }  ][line width=0.75]      (0, 0) circle [x radius= 3.35, y radius= 3.35]   ;
	\draw    (324.34,177) -- (252.66,248.67) ;
	\draw [shift={(251,250.33)}, rotate = 135] [color={rgb, 255:red, 0; green, 0; blue, 0 }  ][line width=0.75]      (0, 0) circle [x radius= 3.35, y radius= 3.35]   ;
	\draw [shift={(326,175.33)}, rotate = 135] [color={rgb, 255:red, 0; green, 0; blue, 0 }  ][line width=0.75]      (0, 0) circle [x radius= 3.35, y radius= 3.35]   ;
	\draw    (327.66,177) -- (399.34,248.67) ;
	\draw [shift={(401,250.33)}, rotate = 45] [color={rgb, 255:red, 0; green, 0; blue, 0 }  ][line width=0.75]      (0, 0) circle [x radius= 3.35, y radius= 3.35]   ;
	\draw [shift={(326,175.33)}, rotate = 45] [color={rgb, 255:red, 0; green, 0; blue, 0 }  ][line width=0.75]      (0, 0) circle [x radius= 3.35, y radius= 3.35]   ;
	\draw    (402.66,248.67) -- (474.34,177) ;
	\draw [shift={(476,175.33)}, rotate = 315] [color={rgb, 255:red, 0; green, 0; blue, 0 }  ][line width=0.75]      (0, 0) circle [x radius= 3.35, y radius= 3.35]   ;
	\draw [shift={(401,250.33)}, rotate = 315] [color={rgb, 255:red, 0; green, 0; blue, 0 }  ][line width=0.75]      (0, 0) circle [x radius= 3.35, y radius= 3.35]   ;
	\draw [line width=1.5]  [dash pattern={on 1.69pt off 2.76pt}]  (150,175.33) -- (173,175.33) ;
	\draw [line width=1.5]  [dash pattern={on 1.69pt off 2.76pt}]  (481,175.33) -- (504,175.33) ;
	\draw [line width=1.5]  [dash pattern={on 1.69pt off 2.76pt}]  (75,250.33) -- (98,250.33) ;
	\draw [line width=1.5]  [dash pattern={on 1.69pt off 2.76pt}]  (406,250.33) -- (429,250.33) ;
	\draw    (177.35,175.33) .. controls (179.02,173.66) and (180.68,173.66) .. (182.35,175.33) .. controls (184.02,177) and (185.68,177) .. (187.35,175.33) .. controls (189.02,173.66) and (190.68,173.66) .. (192.35,175.33) .. controls (194.02,177) and (195.68,177) .. (197.35,175.33) .. controls (199.02,173.66) and (200.68,173.66) .. (202.35,175.33) .. controls (204.02,177) and (205.68,177) .. (207.35,175.33) .. controls (209.02,173.66) and (210.68,173.66) .. (212.35,175.33) .. controls (214.02,177) and (215.68,177) .. (217.35,175.33) .. controls (219.02,173.66) and (220.68,173.66) .. (222.35,175.33) .. controls (224.02,177) and (225.68,177) .. (227.35,175.33) .. controls (229.02,173.66) and (230.68,173.66) .. (232.35,175.33) .. controls (234.02,177) and (235.68,177) .. (237.35,175.33) .. controls (239.02,173.66) and (240.68,173.66) .. (242.35,175.33) .. controls (244.02,177) and (245.68,177) .. (247.35,175.33) .. controls (249.02,173.66) and (250.68,173.66) .. (252.35,175.33) .. controls (254.02,177) and (255.68,177) .. (257.35,175.33) .. controls (259.02,173.66) and (260.68,173.66) .. (262.35,175.33) .. controls (264.02,177) and (265.68,177) .. (267.35,175.33) .. controls (269.02,173.66) and (270.68,173.66) .. (272.35,175.33) .. controls (274.02,177) and (275.68,177) .. (277.35,175.33) .. controls (279.02,173.66) and (280.68,173.66) .. (282.35,175.33) .. controls (284.02,177) and (285.68,177) .. (287.35,175.33) .. controls (289.02,173.66) and (290.68,173.66) .. (292.35,175.33) .. controls (294.02,177) and (295.68,177) .. (297.35,175.33) .. controls (299.02,173.66) and (300.68,173.66) .. (302.35,175.33) .. controls (304.02,177) and (305.68,177) .. (307.35,175.33) .. controls (309.02,173.66) and (310.68,173.66) .. (312.35,175.33) .. controls (314.02,177) and (315.68,177) .. (317.35,175.33) .. controls (319.02,173.66) and (320.68,173.66) .. (322.35,175.33) -- (323.65,175.33) -- (323.65,175.33) ;
	\draw [shift={(326,175.33)}, rotate = 0] [color={rgb, 255:red, 0; green, 0; blue, 0 }  ][line width=0.75]      (0, 0) circle [x radius= 3.35, y radius= 3.35]   ;
	\draw [shift={(175,175.33)}, rotate = 0] [color={rgb, 255:red, 0; green, 0; blue, 0 }  ][line width=0.75]      (0, 0) circle [x radius= 3.35, y radius= 3.35]   ;
	\draw    (328.35,175.33) .. controls (330.02,173.66) and (331.68,173.66) .. (333.35,175.33) .. controls (335.02,177) and (336.68,177) .. (338.35,175.33) .. controls (340.02,173.66) and (341.68,173.66) .. (343.35,175.33) .. controls (345.02,177) and (346.68,177) .. (348.35,175.33) .. controls (350.02,173.66) and (351.68,173.66) .. (353.35,175.33) .. controls (355.02,177) and (356.68,177) .. (358.35,175.33) .. controls (360.02,173.66) and (361.68,173.66) .. (363.35,175.33) .. controls (365.02,177) and (366.68,177) .. (368.35,175.33) .. controls (370.02,173.66) and (371.68,173.66) .. (373.35,175.33) .. controls (375.02,177) and (376.68,177) .. (378.35,175.33) .. controls (380.02,173.66) and (381.68,173.66) .. (383.35,175.33) .. controls (385.02,177) and (386.68,177) .. (388.35,175.33) .. controls (390.02,173.66) and (391.68,173.66) .. (393.35,175.33) .. controls (395.02,177) and (396.68,177) .. (398.35,175.33) .. controls (400.02,173.66) and (401.68,173.66) .. (403.35,175.33) .. controls (405.02,177) and (406.68,177) .. (408.35,175.33) .. controls (410.02,173.66) and (411.68,173.66) .. (413.35,175.33) .. controls (415.02,177) and (416.68,177) .. (418.35,175.33) .. controls (420.02,173.66) and (421.68,173.66) .. (423.35,175.33) .. controls (425.02,177) and (426.68,177) .. (428.35,175.33) .. controls (430.02,173.66) and (431.68,173.66) .. (433.35,175.33) .. controls (435.02,177) and (436.68,177) .. (438.35,175.33) .. controls (440.02,173.66) and (441.68,173.66) .. (443.35,175.33) .. controls (445.02,177) and (446.68,177) .. (448.35,175.33) .. controls (450.02,173.66) and (451.68,173.66) .. (453.35,175.33) .. controls (455.02,177) and (456.68,177) .. (458.35,175.33) .. controls (460.02,173.66) and (461.68,173.66) .. (463.35,175.33) .. controls (465.02,177) and (466.68,177) .. (468.35,175.33) .. controls (470.02,173.66) and (471.68,173.66) .. (473.35,175.33) -- (473.65,175.33) -- (473.65,175.33) ;
	\draw [shift={(476,175.33)}, rotate = 0] [color={rgb, 255:red, 0; green, 0; blue, 0 }  ][line width=0.75]      (0, 0) circle [x radius= 3.35, y radius= 3.35]   ;
	\draw [shift={(326,175.33)}, rotate = 0] [color={rgb, 255:red, 0; green, 0; blue, 0 }  ][line width=0.75]      (0, 0) circle [x radius= 3.35, y radius= 3.35]   ;
	\draw    (102.35,250.33) -- (248.65,250.33) ;
	\draw [shift={(251,250.33)}, rotate = 0] [color={rgb, 255:red, 0; green, 0; blue, 0 }  ][line width=0.75]      (0, 0) circle [x radius= 3.35, y radius= 3.35]   ;
	\draw [shift={(100,250.33)}, rotate = 0] [color={rgb, 255:red, 0; green, 0; blue, 0 }  ][line width=0.75]      (0, 0) circle [x radius= 3.35, y radius= 3.35]   ;
	\draw    (398.65,250.33) -- (253.35,250.33) ;
	\draw [shift={(251,250.33)}, rotate = 180] [color={rgb, 255:red, 0; green, 0; blue, 0 }  ][line width=0.75]      (0, 0) circle [x radius= 3.35, y radius= 3.35]   ;
	\draw [shift={(401,250.33)}, rotate = 180] [color={rgb, 255:red, 0; green, 0; blue, 0 }  ][line width=0.75]      (0, 0) circle [x radius= 3.35, y radius= 3.35]   ;
	
	\draw (226,155.4) node [anchor=north west][inner sep=0.75pt]  [font=\small]  {$r=0$};
	\draw (385,155.4) node [anchor=north west][inner sep=0.75pt]  [font=\small]  {$r=0$};
	\draw (145,252.4) node [anchor=north west][inner sep=0.75pt]  [font=\small]  {$r=\infty $};
	\draw (291,252.4) node [anchor=north west][inner sep=0.75pt]  [font=\small]  {$r=\infty $};
	\draw (191,250.9) node [anchor=north west][inner sep=0.75pt]  [font=\small]  {$\mathscr{I} ^{-}$};
	\draw (338,250.9) node [anchor=north west][inner sep=0.75pt]  [font=\small]  {$\mathscr{I} ^{-}$};
	\draw (368.55,200.24) node [anchor=north west][inner sep=0.75pt]  [font=\small,rotate=-45]  {$r_{c}$};
	\draw (218.55,200.24) node [anchor=north west][inner sep=0.75pt]  [font=\small,rotate=-45]  {$r_{c}$};
	\draw (121.22,209.43) node [anchor=north west][inner sep=0.75pt]  [font=\small,rotate=-315]  {$r_{c}$};
	\draw (421.22,209.43) node [anchor=north west][inner sep=0.75pt]  [font=\small,rotate=-315]  {$r_{c}$};
	\draw (171,155.9) node [anchor=north west][inner sep=0.75pt]  [font=\small]  {$\mathscr{P}$};
	\draw (321,155.9) node [anchor=north west][inner sep=0.75pt]  [font=\small]  {$\mathscr{P}$};
	\draw (471,155.9) node [anchor=north west][inner sep=0.75pt]  [font=\small]  {$\mathscr{P}$};
	\draw (96,254.9) node [anchor=north west][inner sep=0.75pt]  [font=\small]  {$\mathscr{Q}$};
	\draw (246,253.9) node [anchor=north west][inner sep=0.75pt]  [font=\small]  {$\mathscr{Q}$};
	\draw (396,254.9) node [anchor=north west][inner sep=0.75pt]  [font=\small]  {$\mathscr{Q}$};
	\draw (272.22,209.43) node [anchor=north west][inner sep=0.75pt]  [font=\small,rotate=-315]  {$r_{c}$};

\end{tikzpicture}

\caption{Penrose diagram for Nariai-Kiselev BH.}
\label{fig: Penrose Nariai}
\end{figure}
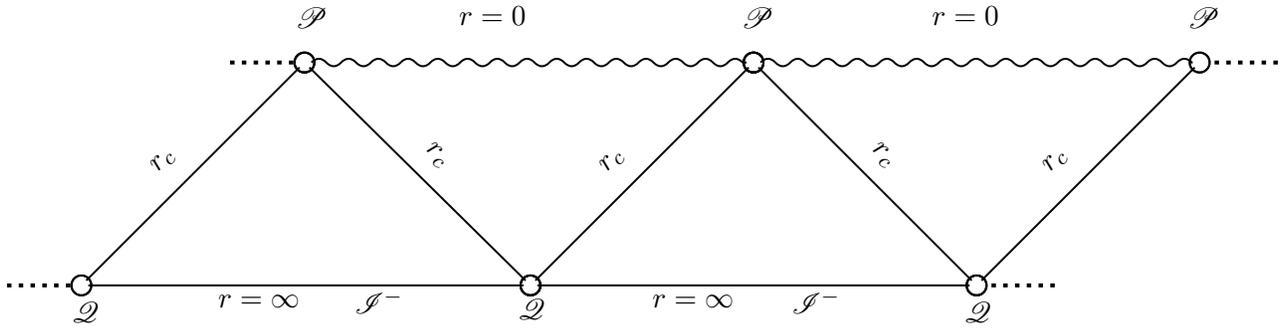

Setting $\omega=-2/3$, the thermodynamical stability of uncharged and charged Kiselev BHs are studied by means of effective thermodynamic quantities in \cite{Ma:2016arz}. It is found that the uncharged Kiselev BH is always thermodynamically unstable due to negative heat capacity, while the charged Kiselev BH will undergo a second-order phase transition. The thermodynamical stability of a BH depends on the sign of the heat capacity which is related to the values of $c$ and $\omega$ in the case of Kiselev BH \cite{Toledo:2018hav}. Moreover, the transition point, in which the heat capacity diverges, changes when we vary $c$ and $\omega$. The phase transition of the rotational Kiselev BH in the case $\omega=1/3$ is calculated in \cite{Xu:2016sew} and it is shown that it is a second-order phase transition. Applying three known approaches: the classical thermodynamical method, the Poincaré method and some geometrical methods, the thermodynamic stability and phase transitions of the asymptotically flat solutions, $-1/3\leq\omega<0$, are investigated in \cite{AzregAinou:2012hy}. It is found that the charged Kiselev BH with low entropy, or high charge or both are local stable. The phase transition of Kiselev BH is also discussed in \cite{Tharanath:2013jt} and the expressions of  mass, density of matter, temperature and heat capacity of the BH is obtained in terms of its entropy. One can see that there is  a discontinuity in the heat capacity of BH which implies that the BH undergoes a phase transition \cite{Tharanath:2013jt}. Using the third-order WKB approximation, the quasinormal frequencies of massless scalar field perturbation around a Kiselev BH is evaluated in \cite{Chen:2005qh}. It is shown that the scalar field damps rapidly and the decay rate would become slow when $\omega$ increases. This study is also done also for gravitational perturbation in \cite{Zhang:2006ij}. We see that the gravitational wave damps slowly and for smaller values of $\omega$, one gets a slower damping.


In the next sections, we are interested in studying a collapsing spherical thin shell,  both null and timelike, in Kiselev space--time.
\section{Null thin shell collapse}\label{Null thin shell collapse}
Here we consider the simplest model of gravitational collapse which is a collapsing thin shell of null matter. It is convenient to use the ingoing Eddington-Finkelstein coordinates which are adopted to the ingoing null geodesics. We assume that the geometry is flat inside the shell and its exterior space--time is described by Kiselev metric. Therefore 
\begin{equation}\label{n1}
ds^2=-f(r)dv^2+2dvdr+r^2d\Omega^2, \quad\quad f(r)= \left\{ 
\begin{tabular}{ll}
$1-\frac{2m}{r}-\frac{c}{r^{3\omega +1}}$ & $v\geq v_0$ \\ 
$1$ & $v< v_0$ \\ 
\end{tabular} 
\right .
\end{equation}  
where $v=t+\int{f(r)^{-1}dr}$. Suppose that the shell moves along the null trajectory $v=v_0$ in both space--times, inside and outside the shell. Therefore
\begin{equation}\label{v}
f(v,r)=1-\left(\frac{2m}{r}-\frac{c}{r^{3\omega +1}}\right)\Theta(v-v_0)
\end{equation} 
in which $\Theta$ is the step function. This is a particular case of Vaidya generalization of Kiselev metric\cite{Heydarzade:2018bli} defined by $m(v)=m\Theta(v-v_0)$ and $c(v)=c\Theta(v-v_0)$. Moreover, the above mentioned metric for $\omega\neq-1$ is a special case of a large family of dynamical BH introduced in \cite{Kothawala:2004fy} \footnote{Setting arbitrary functions and parameters of \cite{Kothawala:2004fy}  as: $M(v)=m\theta(v-v_0)$, $k=-(1+3\omega)/2$ and $C(v)=-3\omega c \theta(v-v_0)/8\pi$, the metric given in (\ref{v}) is resulted.}.  Substituting (\ref{n1}) into Einstein equations gives the following non-vanishing components for energy-momentum tensor
\begin{equation}\label{eq: 1}
T_{vv}=\frac{1}{\kappa}\left(\frac{2m}{r^2}+\frac{c}{r^{3\omega+2}}\right)\delta(v-v_0)-\frac{3c\omega}{\kappa r^{3(\omega+1)}}\left[1-\left(\frac{2m}{r}-\frac{c}{r^{3\omega +1}}\right)\Theta(v-v_0)\right] 
\end{equation}
\begin{equation}\label{eq: 2}
T_{rv}=\frac{3c\omega}{\kappa r^{3(\omega+1)}}\Theta(v-v_0)
\end{equation}
\begin{equation}\label{eq: 3}
T_{\theta\theta}= -\frac{3c\omega(1+3\omega)}{2\kappa r^{3\omega+1}}\Theta(v-v_0),   \hspace{1.5cm}  T_{\phi\phi}= \sin\theta T_{\theta\theta}
\end{equation}
By introducing two future-pointing null vectors $v_{\mu}=(1,0,0,0)$ and \\ $w_{\mu}=\left(g_{vv}/2,-1,0,0\right)$\cite{Husain:1995bf}, one can write the above energy-momentum tensor as
\[
T_{\mu\nu}=\left(\frac{2 m}{\kappa r^2}+\frac{c}{\kappa r^{3\omega+2}}\right)\delta(v-v_0)v_{\mu}v_{\nu}+
\]
\begin{equation} \label{eq: EM temsor dot}
 \Big((\rho+p_t)\left(v_{\mu}w_{\nu}+v_{\nu}w_{\mu}\right)+ p_t g_{\mu\nu}\Big)\Theta(v-v_0)
\end{equation}
It is evident from this relation that the energy flows only along the null direction $w_{\mu}$ since $T_{\mu\nu}v^{\mu}v^{\nu}=0$.
As expected, in the static case, the above energy momentum tensor reduces to
\begin{equation}\label{eq: Kiselev EM2}
T_{\mu\nu}= (\rho+p_t)\left(v_{\mu}w_{\nu}+v_{\nu}w_{\mu}\right)+ p_t g_{\mu\nu}.
\end{equation}
This is the source of Kiselev space--time and as mentioned before, it has not the form of a perfect fluid energy-momentum tensor. 
\section{Timelike thin shell collapse}\label{Timelike thin shell collapse}
Here, we want to consider the gravitational collapse of a timelike spherical thin shell in Kiselev space--time. In contrast to the null case, there is no single coordinate covering both regions, inside and outside the shell and therefore one should introduce two different coordinates. This means that one has to apply the Barrabes-Israel formalism \cite{Barrabes:1991ng} to join two space--times separated by the shell and determine the surface energy-momentum of it.
Below, first we review Barrabes-Israel junction conditions briefly  and then we join outer Kiselev  and inner de Sitter space--times assuming the shell is made of some pressureless matter.
\subsection{Junction Conditions}\label{Junction Conditions}
Let $\Sigma$ be a timelike hypersurface that partitions space--time $\mathcal{V}$ into two parts $\mathcal{V^{\pm}}$. In region $\mathcal{V^{\pm}}$, the metric and coordinate charts are $g^{\pm}_{\alpha\beta}$ and $x^{\alpha}_{\pm}$ respectively. The unit normal vector to $\Sigma$ is $n^{\alpha}$   pointing from $\mathcal{V^{-}}$ to $\mathcal{V^{+}}$ and defined such that
\begin{equation}\label{J1}
n^{\alpha}n_{\alpha}=1, \quad\quad n_{\alpha}e^{\alpha}_{(a)}=0 
\end{equation}
where $e^{\alpha}_{(a)}$ are three basis vectors on $\Sigma$ and have zero jump across $\Sigma$, i.e. \\$[e^{\alpha}_{(a)}]=0$. The first junction condition dictates the continuity of the metric across $\Sigma$: $[g_{\alpha\beta}]=0$. Defining the induced metric on $\Sigma$ as $h_{ab}=g_{\alpha\beta}e^{\alpha}_{(a)}e^{\beta}_{(b)}$, this condition can be written as $[h_{ab}]=0$. The second junction condition relates the energy-momentum tensor of $\Sigma$ to the discontinuity of extrinsic curvature, $K_{ab}$, 
\begin{equation}\label{eq: jump Kab}
\left[K_{ab}\right]=8\pi\left(S_{ab}-\frac{1}{2}h_{ab}S\right)
\end{equation}
where $S_{ab}$ is the energy-momentum of the surface layer $\Sigma$ defined as \\
$T^{\alpha\beta}_{\Sigma}=\delta(\tau) S^{ab}e^{\alpha}_{(a)}e^{\beta}_{(b)}$ and the traces of $K_{ab}$ and $S_{ab}$ are indicated by $K$ and $S$ respectively.

Now, let us find the equation of motion of the shell. To do so, it is straightforward to verify that
the energy momentum conservation equation on the hypersurface reduces to
\begin{equation}\label{eq: Sab conserv}
S^{a}_{b|a} + \left[T_{\alpha\beta}e^{\alpha}_b n^{\beta}\right]=0.
\end{equation}
Here we restrict ourselves to the case that the shell is composed of a pressureless perfect fluid. We will show that such surface energy momentum tensor is required to have a smooth transition across the layer. So, assume
\begin{equation}\label{eq: Sab}
S_{ab}=\sigma u_a u_b
\end{equation}
where $\sigma$ is the surface energy density of the shell and $u^{a}$ is the three-velocity of it. Inserting \eqref{eq: Sab} into \eqref{eq: Sab conserv} leads to
\begin{equation}\label{eq: conserv}
\left(\sigma u^{a}\right)_{|a}=\left[T_{\alpha\beta}u^{\alpha} n^{\beta}\right].
\end{equation} 
The equation of motion of the shell can be found by calculating its acceleration   
\begin{equation}
a^{\alpha}\equiv u^{\alpha}_{;\beta}u^{\beta}= a^{a}e^{\alpha}_{(a)}+u^a u^b K_{ab} n^{\alpha}.
\end{equation}
Projecting it along the layer gives an internal motion of the shell while its normal component, $n_{\alpha}a^{\alpha}=u^a u^b K_{ab}$, describes the motion of the shell. It is also evident that the jump of the normal acceleration, $n_{\alpha}a^{\alpha}$, across $\Sigma$ is related to the jump of  extrinsic curvature. Therefore, making use of \eqref{eq: jump Kab}, we are able to find the shell equation of motion as follows
\begin{equation}\label{eq: eom}
\left[n_{\alpha}a^{\alpha}\right] = 4\pi\sigma.
\end{equation}
In the next section we utilize \eqref{eq: conserv} and \eqref{eq: eom} to investigate a collapsing timelike shell in Kiselev space--time.
\subsection{The motion of a collapsing timelike shell}\label{The motion of a collapsing timelike shell}
Here, we study the collapsing of a timelike thin shell immersed in Kiselev space--time. To do this, we consider that the space--time inside the shell is described by de Sitter geometry. In this way, we can show that Kiselev BH can be matched to de Sitter core by a timelike dust shell and therefore, in principle,  an infinite number of stationary non-singular BH can be constructed. Each of which is labeled by parameter $\omega$. We will return to this point in the next section. 

In order to make things concrete, we will write the metric in both regions as
\begin{equation} \label{eq: f both sides}
ds_{\pm}^2 = -f_{\pm}(r)dt^2 +f_{\pm}^{-1}(r)dr^2+r^2d\Omega^2.
\end{equation}
where $f_{\pm}$ are defined as
\[
f_{-}(r)=1-\frac{\Lambda}{3}r^2,\quad\quad r<R(\tau)
\]
\begin{equation}\label{rbig}
f_{+}(r)=1-\frac{2m}{r}-\frac{c}{r^{3\omega+1}}, \quad\quad r>R(\tau)
\end{equation}
$\Lambda$ is the cosmological constant and  the shell radius is denoted by $R(\tau)$ parameterized by the proper time, $\tau$, of comoving particle on the shell. The line element on $\Sigma$ is then given by
\begin{equation} \label{eq: intrinsict line element}
\left(ds^2\right)_{\Sigma}=-d\tau^2+R(\tau)^2d\Omega^2
\end{equation}
The hypersurface $\Sigma$ is assumed to be timelike throughout the  space--time, i.e. $n^{\alpha}n_{\alpha}=1$. Thus the shell radius must be smaller than the de Sitter horizon $L=\sqrt{\frac{3}{\Lambda}}$. Regarding the region $\mathcal{V^{+}}$, as mentioned before, the positivity of the energy density requires $c\omega<0$. In the case of $\omega>0$, $f_{+}$ blows up at $r\rightarrow0$ and tends to $1$ at enough large values of $r$. This means, either we have a naked singularity, which we have excluded from this study, or we have a BH with at least one horizon, see figure \ref{fig: Horizons}. Therefore, there exists at least an interval of $r$, $0<r<r_-$, where $r_{-}$ is the innermost (Cauchy) horizon radius. In this interval  $f_{+}(r)$ is positive and thus the hypersurface $\Sigma$ is timelike.  For the case that $\omega<0$,  $f_{+}$ tends to minus infinity when $r\rightarrow0$, so the hypersurface $\Sigma$ would be spacelike for $r<r_{-}$. Therefore, here, we only consider the case that $\omega$ is positive. 

According to the first junction condition, the induced metric on both sides of  $\Sigma$ must be the same, $[h_{ab}]=0$. This relation along with equations (\ref{eq: f both sides})-(\ref{eq: intrinsict line element}), gives\footnote{This is equivalent to say that the four-velocity of the shell, $u^{\alpha}=(\dot{t},\dot{R},0,0)$ is a normalized timelike vector,  $u^{\alpha}u_{\alpha}=-1$.} 
\begin{equation}\label{mo1}
\dot{t}=\frac{\beta(R,\dot{R})}{f(R)}
\end{equation}
where $\beta(R,\dot{R})\equiv \sqrt{f(R)+\dot{R}^2}$. It is convenient to choose $e^{\alpha}_{\tau}=u^\alpha$, then from (\ref{J1})
\begin{equation}\label{mo2}
n_{\alpha}=\left(-\dot{R},\frac{\beta}{f(R)},0,0\right)
\end{equation}
By considering  (\ref{mo1}) and (\ref{mo2}), after some straightforward calculations, the non-zero components of extrinsic curvature are derived as follows
\begin{equation}\label{beta}
n_{\alpha}a^{\alpha}=K\indices{^\tau_\tau}=\frac{\dot{\beta}}{\dot{R}},\quad\quad\quad K\indices{^\theta_\theta}=K\indices{^\phi_\phi}=-\frac{\beta}{R}
\end{equation}
Substituting the above relations into the shell equation of motion \eqref{eq: eom}, it can be simplified as
\begin{equation}\label{eq: beta-beta}
\dot{\beta_{+}}-\dot{\beta_{-}}=4\pi \dot{R}\sigma.
\end{equation}

Another useful equation is \eqref{eq: conserv}. By noting (\ref{eq: caconical form}), (\ref{eq: EM final}) and (\ref{eq: f_q}), it can be easily seen that the two terms in the right hand side of \eqref{eq: conserv} are individually zero\footnote{This is because that every component of energy momentum tensor in both regions of space--time is proportional to the corresponding coefficient of metric and also the fact that the velocity and acceleration vectors of the shell are orthogonal.}. Consequently we have 
\begin{equation}\label{eq: R2sigma.}
\left(R^2\sigma\right)^.=0.
\end{equation}
Making use of the two latter equations, one can show that
\begin{equation}\label{eq: eom2}
\beta_{-}-\beta_{+}=\frac{M}{R}+ const.
\end{equation}
where $M\equiv 4\pi R^2 \sigma$ is the proper shell mass which is constant by virtue of equation \eqref{eq: R2sigma.}. Also the constant of (\ref{eq: eom2}) is equal to zero. This can be easily verified by substituting (\ref{eq: jump Kab}) and (\ref{eq: Sab}) into the second relation of (\ref{beta}).
 
Now, a question may be raised here. Is it possible to have a stable stationary shell by adjusting the free parameters of shell and geometry?
This is the subject of the next section.
\subsection{Stable regular BH}\label{secRBH}
In this section, we have found some appropriate ranges for shell radius and its mass and also for three parameters of Kiselev metric ($m$, $c$, $\omega$) for which a stable stationary BH is constructed. To do this, by aid of (\ref{rbig}), we insert the definition of $\beta$  into (\ref{eq: eom2}).
This reduces (\ref{eq: eom2}) in the form of a conservation law
\begin{equation}\label{eq: eom3}
	\dot{R^2} + V(R)=-1
\end{equation}
where
\begin{equation}\label{eq: V(R)}
	V(R)=-\left[\frac{-2m-\frac{c}{R^{3\omega}}+\frac{R^3}{L^2}}{2M}-\frac{M}{2R}\right]^2-\frac{R^2}{L^2}.
\end{equation}
is the effective potential of shell. For stationary BHs, $\dot{R}=0$, and so $V(R)=-1$ and $dV(R)/dR=0$ and the stability of solution will be guaranteed by the constraint that the sign of $d^2 V(R)/d R^2$ should be positive. 

Here, we perform a numerical analysis of (\ref{eq: eom3}) and (\ref{eq: V(R)}) to get more insight regarding a stable regular Kiselev BH. Without loss of generality, we set $L=1$ and normalize other parameters as follows: $R/L\rightarrow R$, $m/L\rightarrow m$, $M/L\rightarrow M$ and $c^{\frac{1}{3\omega+1}}/L\rightarrow c^{\frac{1}{3\omega+1}}$ to get dimensionless parameters. Solving $V(R)+1=0$ and $dV(R)/dR=0$ simultaneously,  gives us the following relations for $m$ and $c$ in terms of $M$, $R$ and $\omega$ \footnote{There is also another set of relations for $m$ and $c$,  for which \eqref{eq: eom2} is only satisfied if  $M=0$, therefore it does not lead to a valid solution.} 
\begin{equation}\label{eq: set1 a}	m(M,R,\omega)=\frac{M^2(1-3 \omega)+3 R^4(\omega +1)}{6R\omega}-\frac{M\left(3 \left(R^2-1\right)\omega+R^2\right)}{3\sqrt{1-R^2}\omega}
\end{equation}
\begin{equation}\label{eq: set1 b}
c(M,R,\omega)=\frac{R^{3\omega-1}}{3\omega}\left(\frac{2MR^3}{\sqrt{1-R^2}}-M^2-3R^4\right)
\end{equation}
Inserting \eqref{eq: set1 a} and \eqref{eq: set1 b} into \eqref{eq: eom2} yields
\begin{equation}
	M+\abs{M-R\sqrt{1-R^2}}=R\sqrt{1-R^2}.
\end{equation}
which is satisfied if $M<R\sqrt{1-R^2}$. Since the normalized radius of the shell belongs to $0<R<1$, we find that $M<1/2$. Putting these together and noting that $\omega>0$, we find that \eqref{eq: set1 a} and \eqref{eq: set1 b}  lead to the stationary solutions as long as $m>0$. In addition, the resulted solutions will be stable if they satisfy  $d^2V(R)/dR^2>0$. Moreover, we assume that there is no naked singularity. This assumption can be expressed as $\tilde{c}\geq\tilde{c}_{ext}$ derived earlier in section \ref{Horizon}. This condition strongly affects the acceptable range of the shell radius and therefore its mass. Also, as mentioned before, the shell radius must be smaller than the innermost Kiselev horizon to have a timelike shell. These constraints can be shown diagrammatically. Figure \ref{fig: mwr} illustrates the allowed regions of parameter space $(M, R, \omega)$, $(m, R, \omega)$ and $(c, R, \omega)$ by taking into account all conditions explained above. This figure indicates that  there are stable solutions with both negative and positive values of $M$.

\begin{figure}[H]
	\centering
	\subfigure[]{\includegraphics[width=0.37\textwidth]{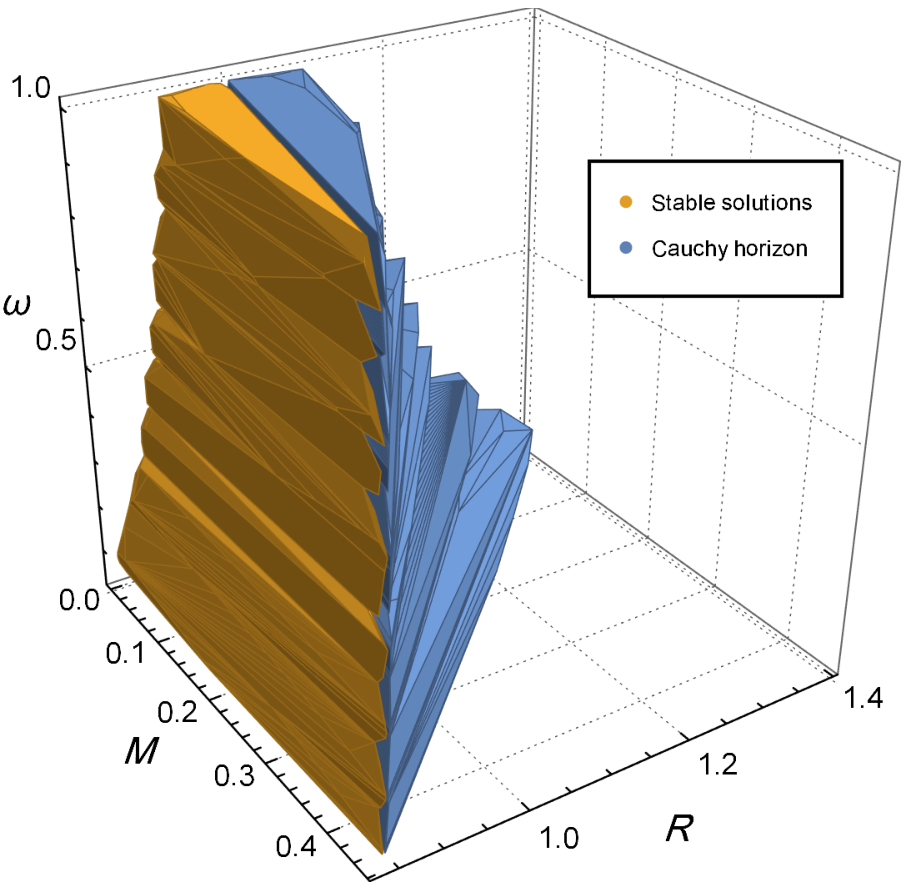}}
	\subfigure[]{\includegraphics[width=0.37\textwidth]{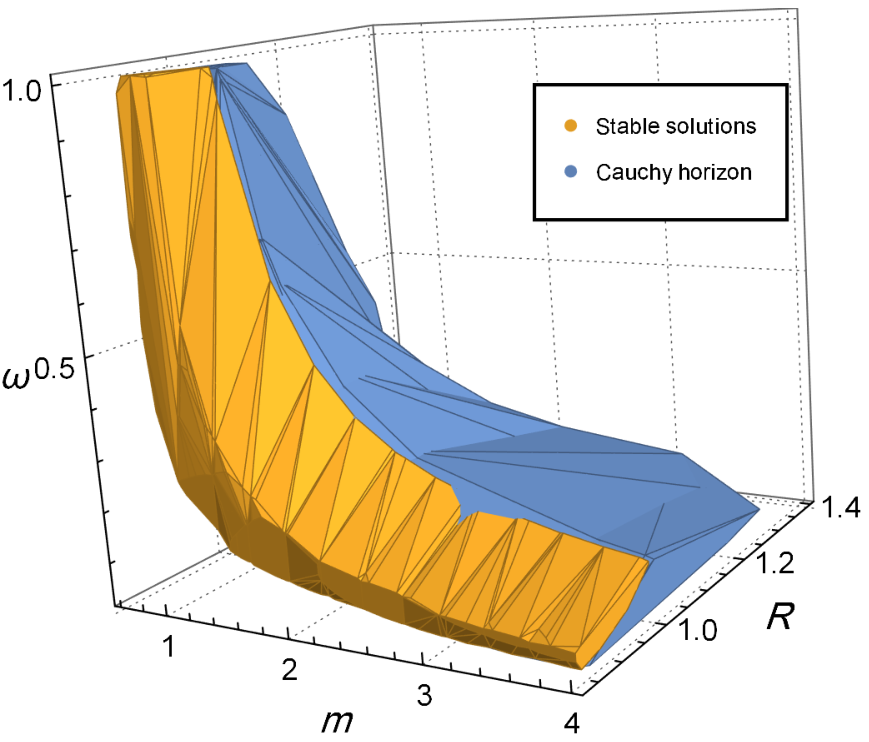}}
	
	\subfigure[]{\includegraphics[width=0.37\textwidth]{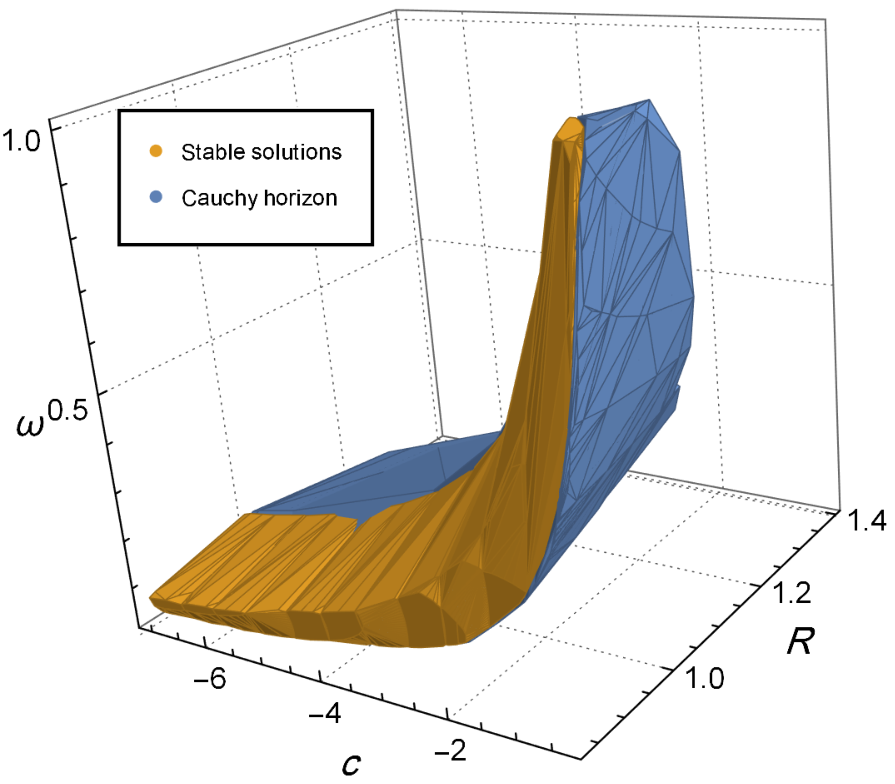}}
	\subfigure[]{\includegraphics[width=0.37\textwidth]{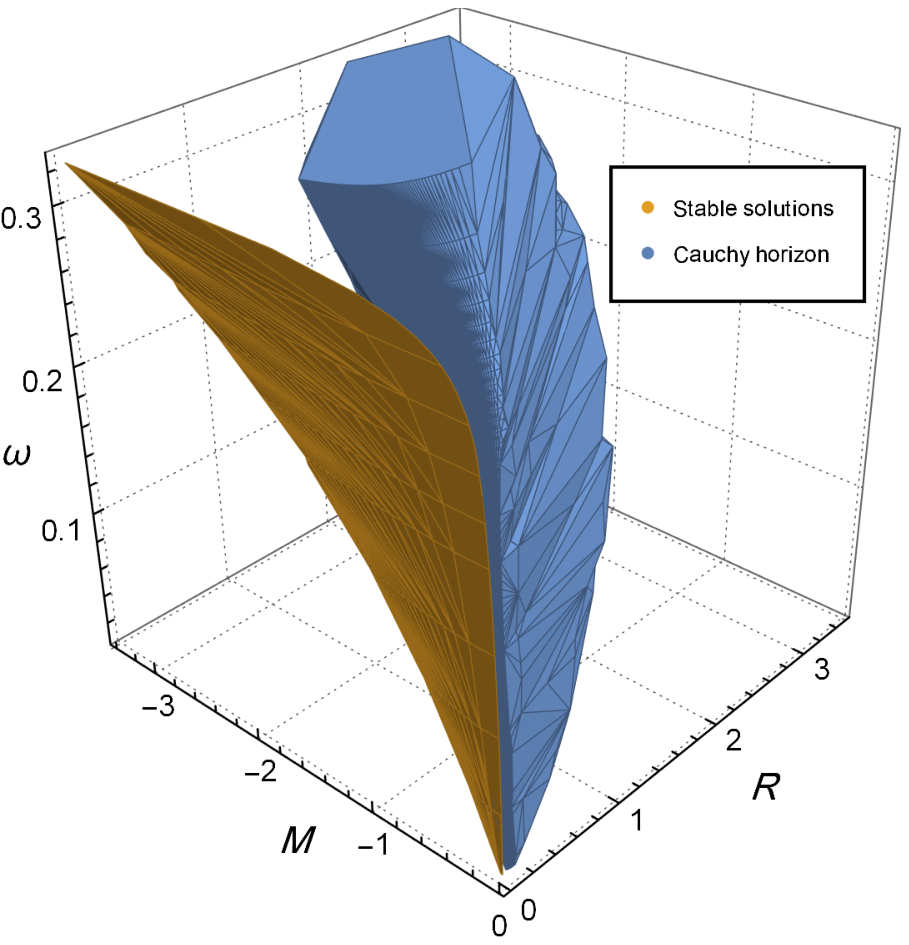}}

	\subfigure[]{\includegraphics[width=0.37\textwidth]{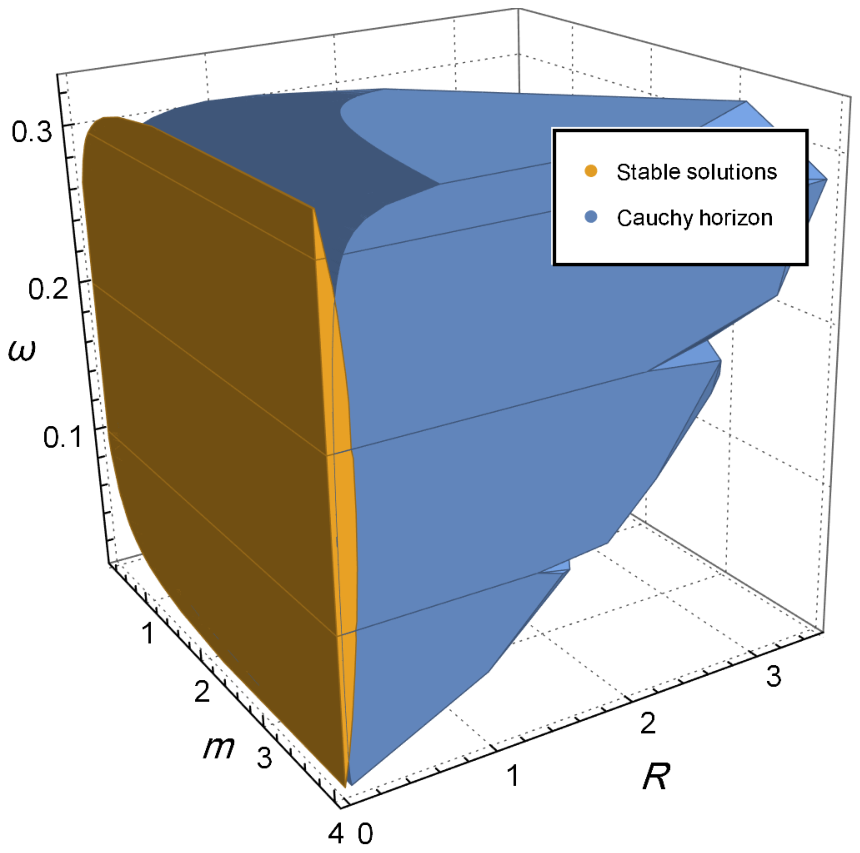}}
	\subfigure[]{\includegraphics[width=0.37\textwidth]{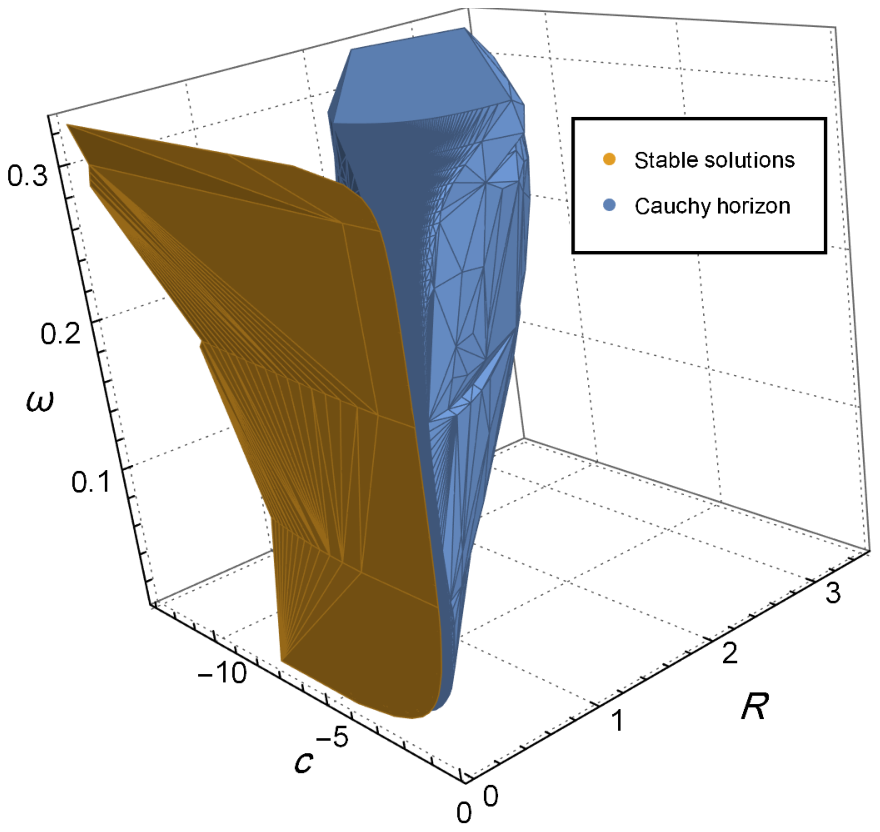}}
	\caption{The valid ranges of parameters (yellow regions), and Cauchy horizon location (blue region) for stable stationary BH solutions. (a), (b), and (c) correspond to solutions with positive shell's mass $M$ and (d), (e), and (f) correspond to negative ones.}
	\label{fig: mwr}
\end{figure}
Furthermore, the allowed ranges of $M$, $m$ and $c$ are unbounded for $\omega<1/3$ for which the negative shell mass can be occurred. This motivates us to consider the special case $\omega=1/3$ in the following in more detail. 

For $\omega=1/3$, the relations \eqref{eq: set1 a} and \eqref{eq: set1 b} read
\begin{equation}\label{eq: rei shell mass}
	M=\frac{\sqrt{1-R^2} \left(2 R^3-m\right)}{2 R^2-1}.
\end{equation}
\begin{equation}\label{eq: rei c}
	c=\frac{m^2 \left(R^2-1\right)+m \left(6 R^3-8 R^5\right)+\left(4 R^2-3\right) R^4}{\left(1-2 R^2\right)^2}.
\end{equation} 
Then, making use of \eqref{eq: rei c}, we are able to find the horizons of Kiselev BH, (\ref{eq: f_q}), as a function of its mass and the shell radius as follows
\begin{equation}
	r_{\pm}= m\pm R \frac{\left|m-R\right|\sqrt{4R^2-3}}{\left|1-2R^2\right|}.
\end{equation}
Therefore, the shell radius must satisfy $R\geq\frac{\sqrt{3}}{2}$. Further examinations reveal that the hypersurface $\Sigma$ remains timelike only if 
\begin{equation}\label{lim1}
	m\geq R\geq \sqrt{3}/2,\hspace{.2cm} m\neq\sqrt{3}/2 
\end{equation}
The upper limit of the BH mass can be deduced from positivity condition of $d^2 V(R)/dR^2$
\begin{equation}\label{eq: Rei Potential}
	2 \left(4+\frac{1}{R^2-1}+\frac{6 (R-m)}{m-2 R^3}\right)>0
\end{equation}
The inequalities (\ref{lim1}) and (\ref{eq: Rei Potential}) give $\frac{\sqrt{3}}{2}<m<2$ and $\frac{\sqrt{3}}{2}<R<1$. Moreover using \eqref{eq: rei shell mass} and \eqref{eq: rei c}, we find that $0<M<\frac{\sqrt{3}}{4}$ and $-3<c<-3/4$. 
\begin{figure}[H]
	\centering
	\subfigure[]{\includegraphics[width=0.49\textwidth]{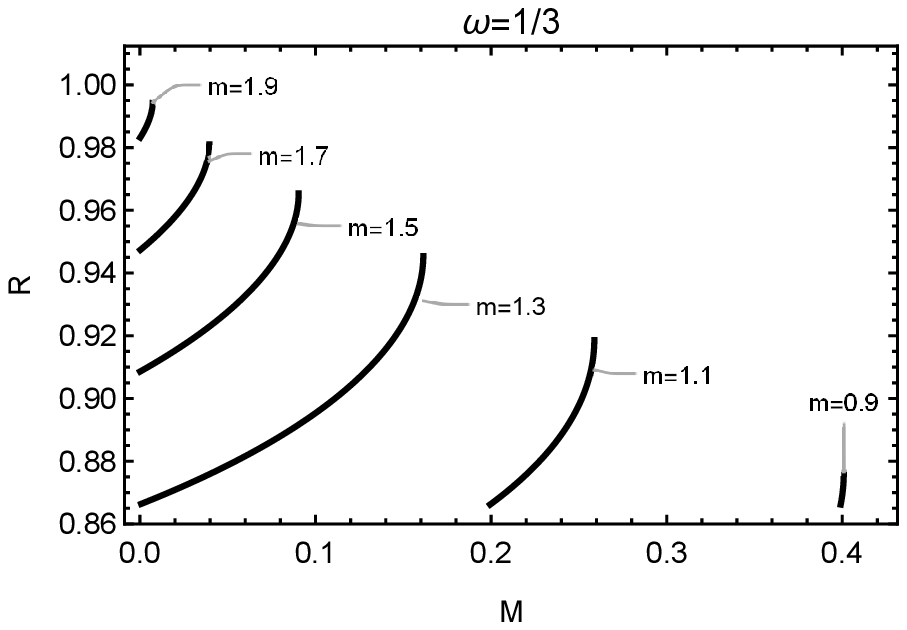}}
	\subfigure[]{\includegraphics[width=0.49\textwidth]{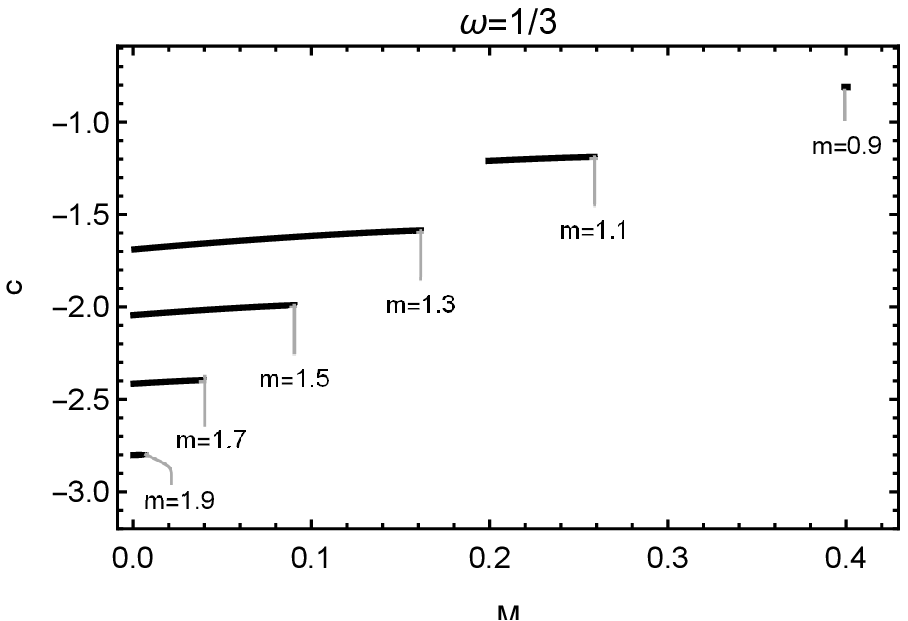}}
	\caption  {Shell radius (a) and the parameter $c$ (b) as functions of shell's mass for different BH mass with $\omega=1/3$.}
	\label{fig: rei}
\end{figure}
In figure \ref{fig: rei}(a) and figure \ref{fig: rei}(b), we have plotted $R(M)$ and $c(M)$ for stable stationary BH solutions considering different values of $m$. We see that as $m$ increases, the maximum values of $R$ and $\abs{c}$ also increase whereas the maximum of $M$ decreases. For a given $m$, it is also evident that by increasing $M$, the shell radius $R$ increases but $\abs{c}$ decreases. 

Now let us turn back to the general case in which $\omega$ has an arbitrary value. In this case,  the coefficient of $M^2$ in \eqref{eq: set1 a} is nonzero and therefore an analytical analysis is impossible. Consequently, we confine ourselves to the numerical analysis in this case. To obtain some sequences of stable stationary regular BHs with fixed values of $m$ and $\omega$,  the shell mass, $M(R)$, is found from equation \eqref{eq: set1 a}. Once $M(R)$ is determined, the parameter $c$ can be obtained from \eqref{eq: set1 b}. As it was mentioned before, for $\omega<1/3$, there are stable BH solutions with both positive and negative shell mass. Therefore, in some interval of $m$, the plot of $R(M)$ and $c(M)$ is expected to be discontinuous for $\omega<1/3$. To see this, $R(M)$ and $c(M)$ are plotted in \ref{fig: lesser 1/3RM}(a), \ref{fig: lesser 1/3RM}(c) with $\omega=0.1$ and \ref{fig: lesser 1/3RM}(b), \ref{fig: lesser 1/3RM}(d) with $\omega=0.2$ for different values of $m$. It is evident  that for $1.9\lessapprox m\lessapprox5.4$ ($\omega=0.1$) and $1.2\lessapprox m\lessapprox2.9$ ($\omega=0.2$), two sets of solutions corresponding to different signs of $M$ are resulted. For example, setting $\omega=0.1$ and $m=2.3$, the normalized shell mass might be found on $-0.42\lesssim M <0$ or $0.28\lesssim M \lesssim0.32$ for stable BHs. In these intervals, let us consider two stable stationary regular solutions as $M\approx0.310$, $R= 0.864$, $c\approx -3.592$ and $M\approx-0.310$, $R= 0.043$, $c\approx -2.900$. For each of these solutions, the effective potential is plotted in figures \ref{fig: degen}(a) and \ref{fig: degen}(b) respectively. In these figures, the local minimum satisfies all conditions mentioned at the beginning of this section for a stable stationary regular BH solution. Also from figure \ref{fig: lesser 1/3RM}, it turns out that with fixed values of $\omega$ and $m$, by increasing $\abs{M}$, the shell remains stable if $R$ and also $\abs{c}$ increases. Moreover we see that for a specific value of $\omega$, increasing $m$ leads to a larger upper limit of $R$ and $\abs{c}$ and also increasing (decreasing) the maximum of $\abs{M}$ in the region $M<0$ ($M>0$).

Assuming $M>0$, it can be seen from figure \ref{fig: mwr} that the absolute values of $m$ and $c$ diverge as $R\rightarrow0$. Moreover, for a stable solution, there is a minimum value for the shell's radius. It is resulted from the existence of cosmological constant inside the shell and corresponds with small values of $\omega$. As an example, for $\omega=10^{-4}$, the local minimum  occurs at  $R_{min}\approx0.806$. A general physical interpretation can be obtained by noting that at final stages of gravitational collapse, the quantum effects can prevent the formation of singularity. As mentioned before, one can consider that there is an upper limit of the order of Planck scale for the curvature of space--time \cite{Frolov:1989pf} or  the de Sitter horizon is of the order of Planck scale and it is much smaller than the event horizon \cite{Balbinot:1990zz}. In both cases, the resisting source against continuation of the collapse is the quantum effects.
\begin{figure}[H]
	\centering
	\subfigure[]{\includegraphics[width=0.49\textwidth]{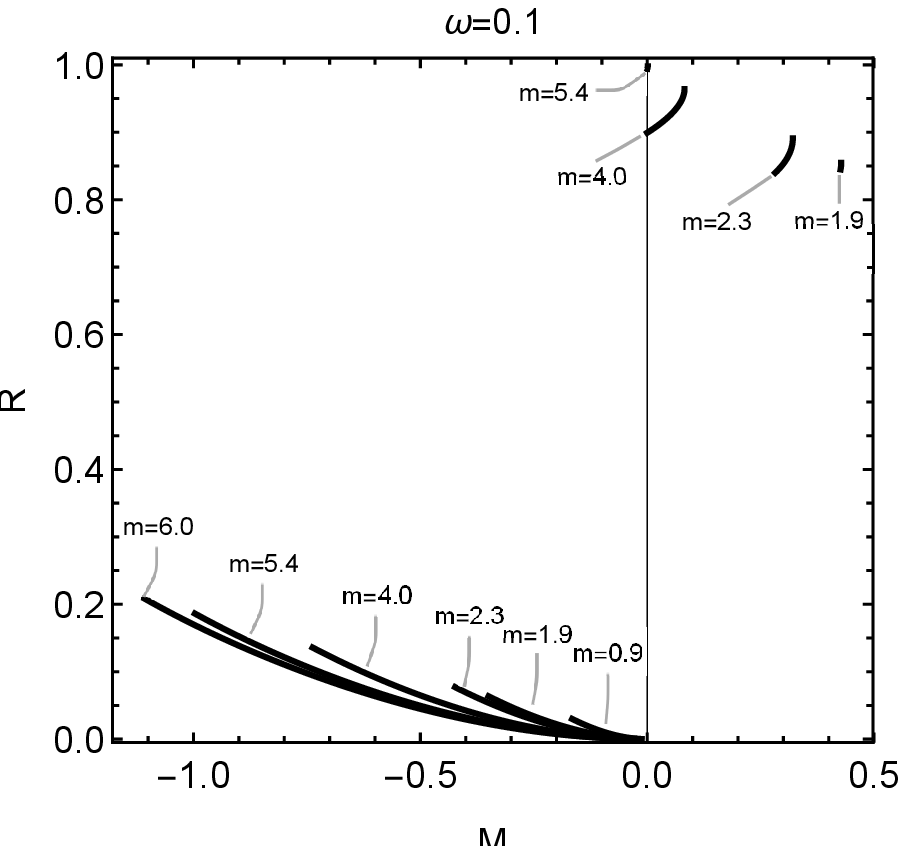}}
	\subfigure[]{\includegraphics[width=0.49\textwidth]{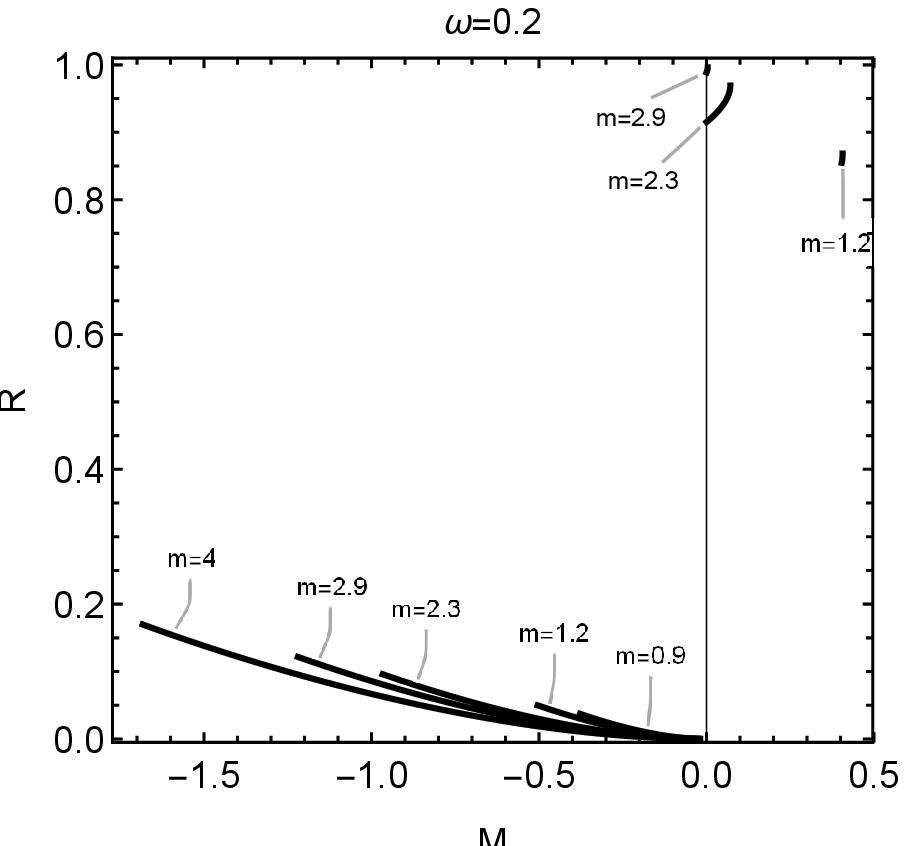}}
	\subfigure[]{\includegraphics[width=0.49\textwidth]{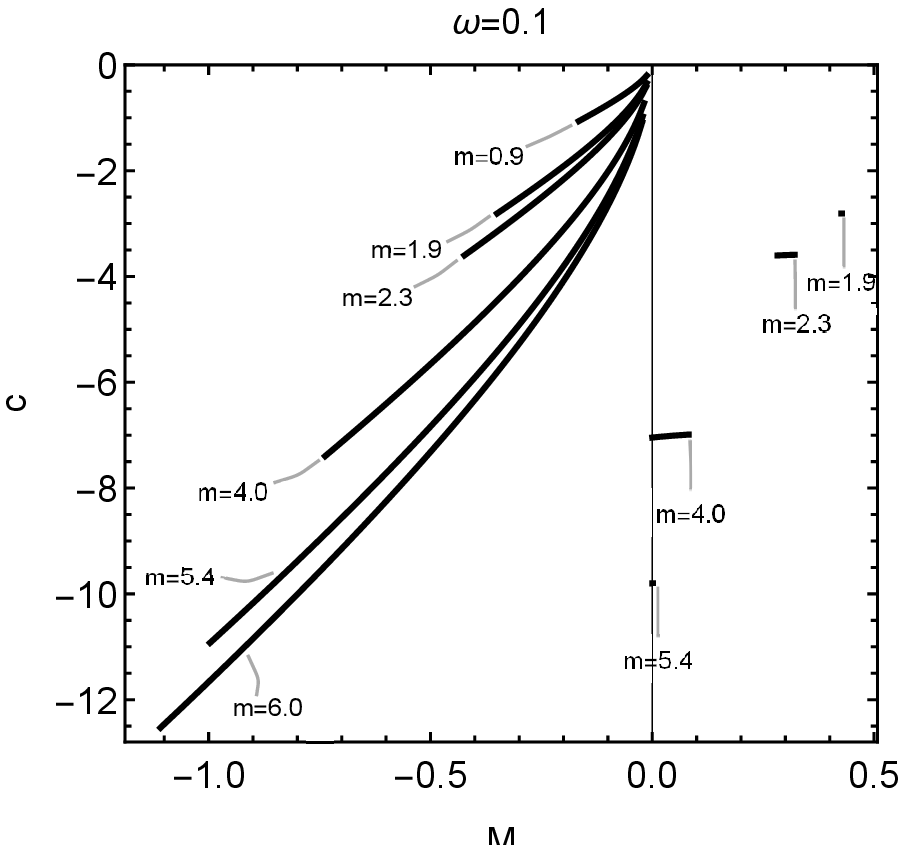}}
	\subfigure[]{\includegraphics[width=0.49\textwidth]{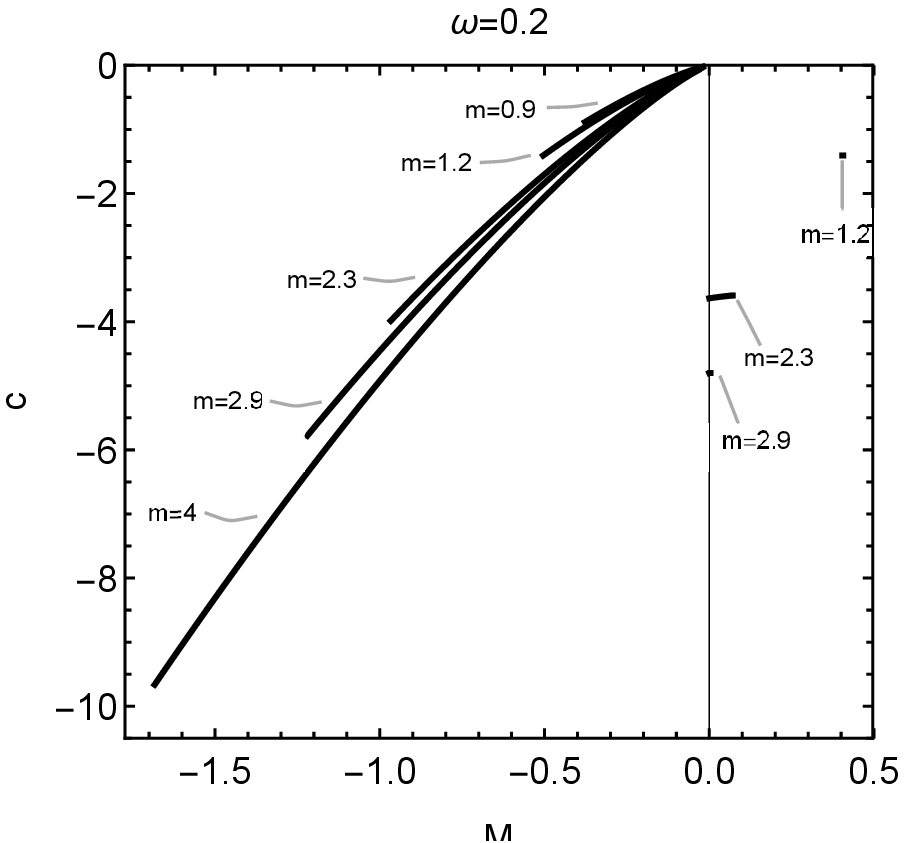}}
	\caption{$R(M)$ and $c(M)$ for stable stationary regular BHs with $\omega=0.1$ in (a) and (c), $\omega=0.2$ in (b) and (d).} 
	\label{fig: lesser 1/3RM}
\end{figure}
\begin{figure}[H]
	\centering
	\subfigure[]{\includegraphics[width=0.6\textwidth]{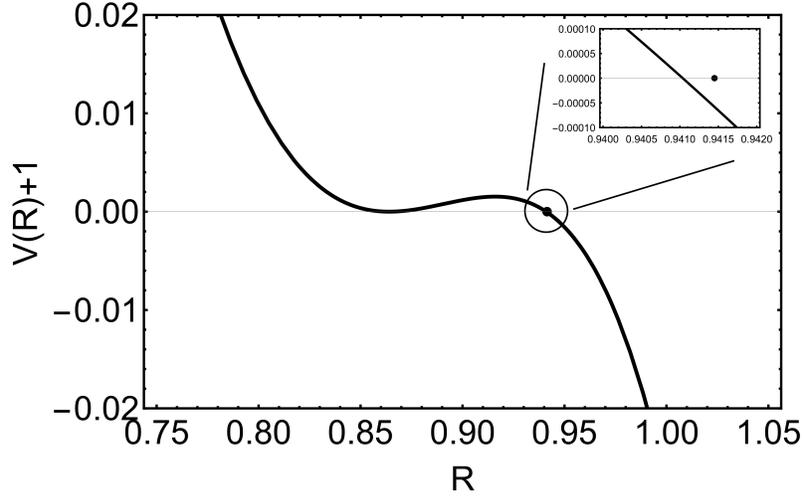}}
	
	\subfigure[]{\includegraphics[width=0.6\textwidth]{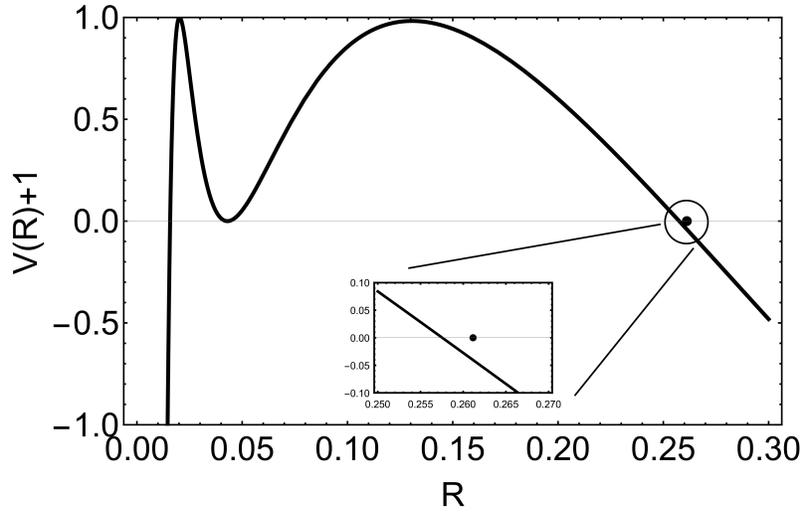}}
	\caption{The shell effective potential  with $\omega=0.1$, and $m=2.3$. (a) First solution with $M\approx0.310$ $c\approx -3.592$. (b) Second solution with $M\approx-0.310$, $c\approx -2.900$. Dots indicate the location of Cauchy horizon.} 
	\label{fig: degen}
\end{figure}
The functions $R(M)$ and $c(M)$ for stable stationary BH solutions with $\omega=0.5$ and $\omega=0.8$ are plotted  in figure \ref{fig: 0.5 and 0.8}(a,b) and \ref{fig: 0.5 and 0.8}(c,d) respectively.  Again we see that with fixed values of $\omega$ and $m$, increasing $M$, leads to increasing $R$ and decreasing $|c|$ for a stable shell. Also, with a fixed value of $\omega$, the maximum values of $R$ and $\abs{c}$ increase by increasing $m$ whereas the maximum value of $M$ decreases.

\begin{figure}[H]
	\centering
	\subfigure[]{\includegraphics[width=0.48\textwidth]{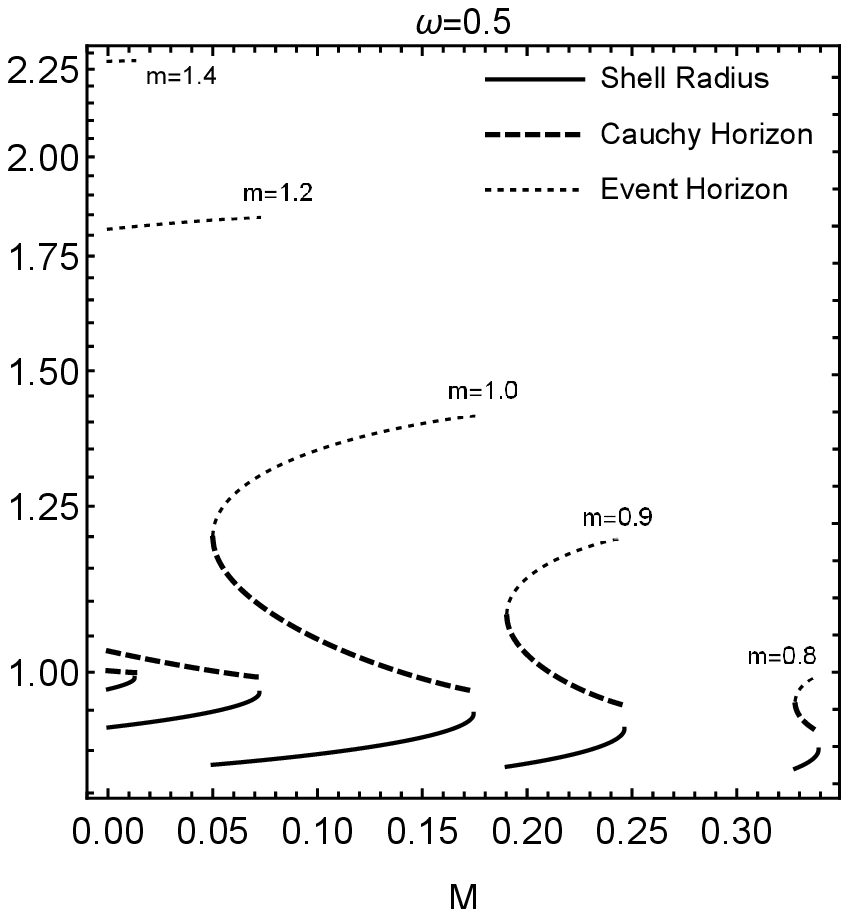}}
	\subfigure[]{\includegraphics[width=0.48\textwidth]{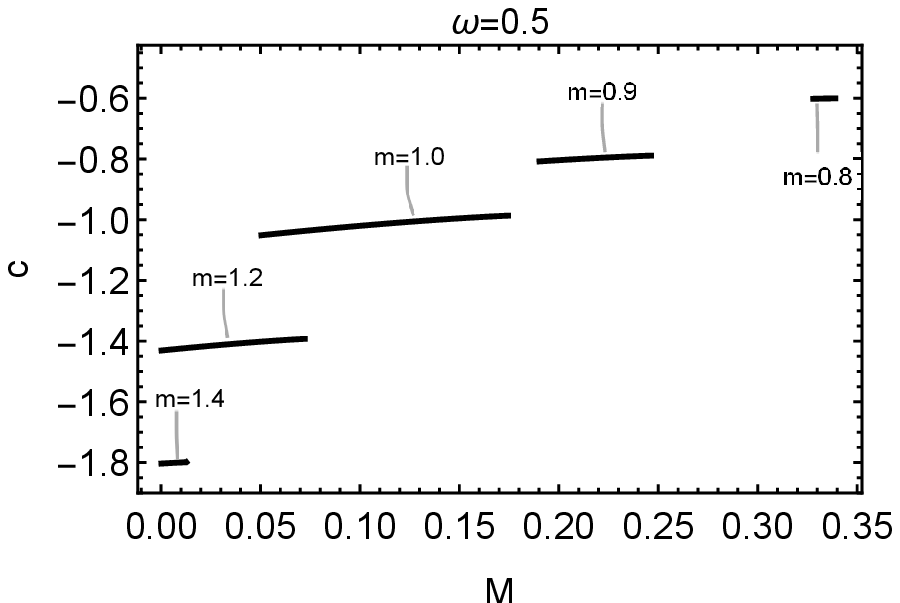}}
	
	\subfigure[]{\includegraphics[width=0.48\textwidth]{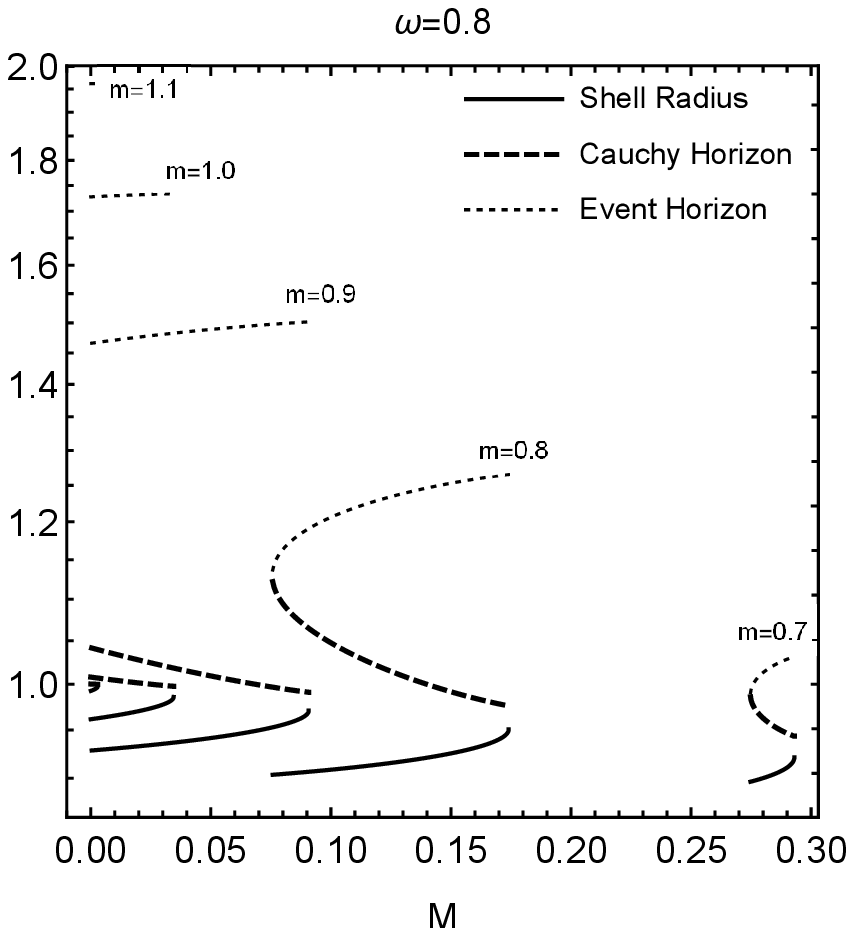}}
	\subfigure[]{\includegraphics[width=0.48\textwidth]{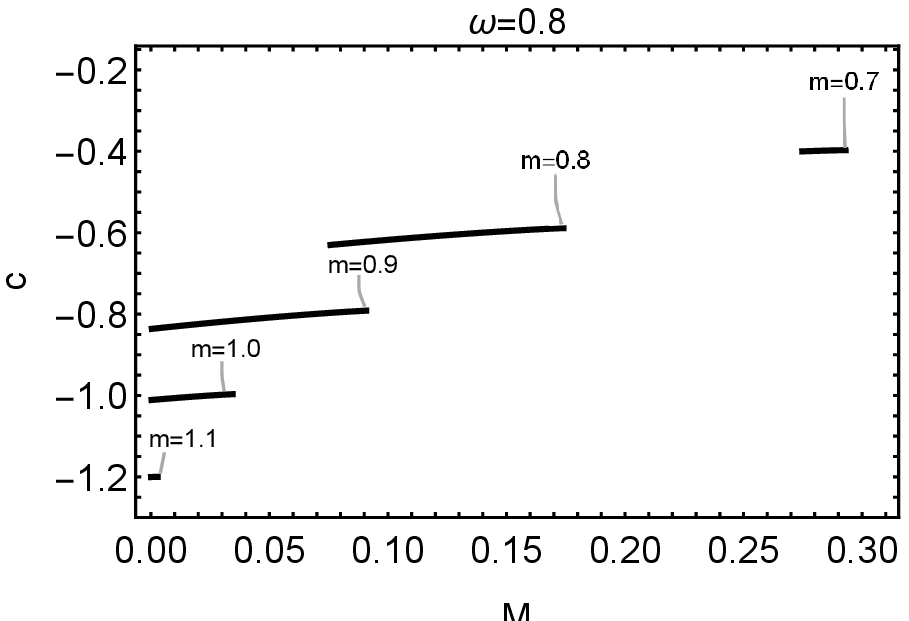}}
	\caption{$R(M)$ (black), event horizon location (blue) and Cauchy horizon location (red) and $c(M)$. $\omega=0.5$ for (a) and (b) and $\omega=0.8$ for (c) and (d).} 
	\label{fig: 0.5 and 0.8}
\end{figure}
The Penrose diagram of the stable thin shell can be obtained when de Sitter space--time is
attached to the Kiselev one through a timelike thin shell inside the Kiselev Cauchy horizon, see figure \ref{fig: Penrose stable}. In this case  the Kiselev singularity at $r=0$ is replaced with de Sitter core.
 
\begin{figure}[H]
\centering
\tikzset{every picture/.style={line width=0.75pt}} 

\begin{tikzpicture}[x=0.75pt,y=0.75pt,yscale=-1.5,xscale=1.5]
	
	\draw    (25,250) -- (75,300) ;
	\draw    (175,300) -- (225,250) ;
	\draw    (175,200) -- (225,250) ;
	\draw [line width=0.75]  [dash pattern={on 0.84pt off 2.51pt}]  (75,100) -- (175,200) ;
	\draw    (75,200) -- (25,250) ;
	\draw [line width=0.75]  [dash pattern={on 0.84pt off 2.51pt}]  (175,100) -- (75,200) ;
	\draw [line width=0.75]  [dash pattern={on 0.84pt off 2.51pt}]  (75,200) -- (175,300) ;
	\draw [line width=0.75]  [dash pattern={on 0.84pt off 2.51pt}]  (75,300) -- (175,200) ;
	\draw [line width=0.75]  [dash pattern={on 3pt off 5pt}][color=black  ,draw opacity=1 ][fill={rgb, 255:red, 0; green, 0; blue, 0 }  ,fill opacity=0.05 ]   (175,100) .. controls (164,142.33) and (164,149.33) .. (175,200) ;	
	\draw    (175,100) -- (175,200) ;
	\draw [fill={rgb, 255:red, 0; green, 0; blue, 0 }  ,fill opacity=0.2 ]   (75,100) -- (75,200) ;
	\draw    (175,300) -- (175,400) ;
	\draw    (75,300) -- (75,400) ;
	\draw [line width=0.75]  [dash pattern={on 0.84pt off 2.51pt}]  (75,300) -- (175,400) ;
	\draw [line width=0.75]  [dash pattern={on 0.84pt off 2.51pt}]  (75,400) -- (175,300) ;
	\draw    (175,100) -- (225,50) ;
	\draw    (25,50) -- (75,100) ;
	\draw [line width=0.75]  [dash pattern={on 0.84pt off 2.51pt}]  (75.43,100.06) -- (125,50) ;
	\draw [line width=0.75]  [dash pattern={on 0.84pt off 2.51pt}]  (125,50) -- (175,100) ;
	\draw    (175,400) -- (225,450) ;
	\draw    (75,400) -- (25,450) ;
	\draw [line width=0.75]  [dash pattern={on 0.84pt off 2.51pt}]  (125,450) -- (175,400) ;
	\draw [line width=0.75]  [dash pattern={on 0.84pt off 2.51pt}]  (75,400) -- (125,450) ;
	\draw [line width=0.75]  [dash pattern={on 3pt off 5pt}][color=black  ,draw opacity=1 ][fill={rgb, 255:red, 0; green, 0; blue, 0 }  ,fill opacity=0.05 ]   (75.43,200.06) .. controls (86.43,157.72) and (86.43,150.72) .. (75.43,100.06) ;
	\draw [line width=0.75]  [dash pattern={on 3pt off 5pt}][color=black  ,draw opacity=1 ][fill={rgb, 255:red, 0; green, 0; blue, 0 }  ,fill opacity=0.05 ]   (175,300) .. controls (164,342.33) and (164,349.33) .. (175,400) ;
	\draw [line width=0.75]  [dash pattern={on 3pt off 5pt}][color=black  ,draw opacity=1 ][fill={rgb, 255:red, 0; green, 0; blue, 0 }  ,fill opacity=0.05 ]   (75,400) .. controls (86,357.67) and (86,350.67) .. (75,300) ;
	\draw   (265,226) -- (380,226) -- (380,276) -- (265,276) -- cycle ;
	\draw [line width=0.75]  [dash pattern={on 3pt off 5pt}][color=black  ,draw opacity=1 ][line width=1.5]    (270,237) -- (295,237) ;
	\draw  [color={rgb, 255:red, 0; green, 0; blue, 0 }  ,draw opacity=0.05 ][fill={rgb, 255:red, 0; green, 0; blue, 0 }  ,fill opacity=0.1 ] (270.5,258) -- (295,258) -- (295,268) -- (270.5,268) -- cycle ;

	\draw (178,141.4) node [anchor=north west][inner sep=0.75pt]  [font=\footnotesize]  {$r=0$};
	\draw (41,141.4) node [anchor=north west][inner sep=0.75pt]  [font=\footnotesize]  {$r=0$};
	\draw (230,241.4) node [anchor=north west][inner sep=0.75pt]  [font=\footnotesize]  {$i^{0}$};
	\draw (12,241.4) node [anchor=north west][inner sep=0.75pt]  [font=\footnotesize]  {$i^{0}$};
	\draw (198,275.4) node [anchor=north west][inner sep=0.75pt]  [font=\footnotesize]  {$\mathscr{I}^{-}$};
	\draw (33,280.4) node [anchor=north west][inner sep=0.75pt]  [font=\footnotesize]  {$\mathscr{I}^{-}$};
	\draw (200,212.4) node [anchor=north west][inner sep=0.75pt]  [font=\footnotesize]  {$\mathscr{I}^{+}$};
	\draw (33,206.4) node [anchor=north west][inner sep=0.75pt]  [font=\footnotesize]  {$\mathscr{I}^{+}$};
	\draw (158.24,272.71) node  [font=\scriptsize,rotate=-45]  {$r=r_{+}$};
	\draw (106.59,221.06) node  [font=\scriptsize,rotate=-45]  {$r=r_{+}$};
	\draw (107.59,421.06) node  [font=\scriptsize,rotate=-45]  {$r=r_{+}$};
	\draw (128.92,427.95) node [anchor=north west][inner sep=0.75pt]  [font=\scriptsize,rotate=-315]  {$r=r_{+}$};
	\draw (129.05,229.74) node [anchor=north west][inner sep=0.75pt]  [font=\scriptsize,rotate=-315]  {$r=r_{+}$};
	\draw (81.4,274.38) node [anchor=north west][inner sep=0.75pt]  [font=\scriptsize,rotate=-315]  {$r=r_{+}$};
	\draw (153.24,168.32) node  [font=\scriptsize,rotate=-45]  {$r=r_{-}$};
	\draw (106.59,120.67) node  [font=\scriptsize,rotate=-45]  {$r=r_{-}$};
	\draw (129.05,129.35) node [anchor=north west][inner sep=0.75pt]  [font=\scriptsize,rotate=-315]  {$r=r_{-}$};
	\draw (85.4,170.99) node [anchor=north west][inner sep=0.75pt]  [font=\scriptsize,rotate=-315]  {$r=r_{-}$};
	\draw (158.24,72.32) node  [font=\scriptsize,rotate=-45]  {$r=r_{+}$};
	\draw (81.4,73.99) node [anchor=north west][inner sep=0.75pt]  [font=\scriptsize,rotate=-315]  {$r=r_{+}$};
	\draw (153.24,367.68) node  [font=\scriptsize,rotate=-45]  {$r=r_{-}$};
	\draw (106.59,321.03) node  [font=\scriptsize,rotate=-45]  {$r=r_{-}$};
	\draw (129.05,329.7) node [anchor=north west][inner sep=0.75pt]  [font=\scriptsize,rotate=-315]  {$r=r_{-}$};
	\draw (84.4,371.35) node [anchor=north west][inner sep=0.75pt]  [font=\scriptsize,rotate=-315]  {$r=r_{-}$};
	\draw (197,75.4) node [anchor=north west][inner sep=0.75pt]  [font=\footnotesize]  {$\mathscr{I}^{-}$};
	\draw (34,81.4) node [anchor=north west][inner sep=0.75pt]  [font=\footnotesize]  {$\mathscr{I}^{-}$};
	\draw (200,412.4) node [anchor=north west][inner sep=0.75pt]  [font=\footnotesize]  {$\mathscr{I}^{+}$};
	\draw (34,405.4) node [anchor=north west][inner sep=0.75pt]  [font=\footnotesize]  {$\mathscr{I}^{+}$};
	\draw (178,342.4) node [anchor=north west][inner sep=0.75pt]  [font=\footnotesize]  {$r=0$};
	\draw (41,342.4) node [anchor=north west][inner sep=0.75pt]  [font=\footnotesize]  {$r=0$};
	\draw (305,231) node [anchor=north west][inner sep=0.75pt]  [font=\footnotesize] [align=left] {Shell radius};
	\draw (301,256) node [anchor=north west][inner sep=0.75pt]  [font=\footnotesize] [align=left] {de Sitter core};

\end{tikzpicture}

\caption{Penrose diagram for stable thin shell solutions with typical values of parameters.}
\label{fig: Penrose stable}
\end{figure}
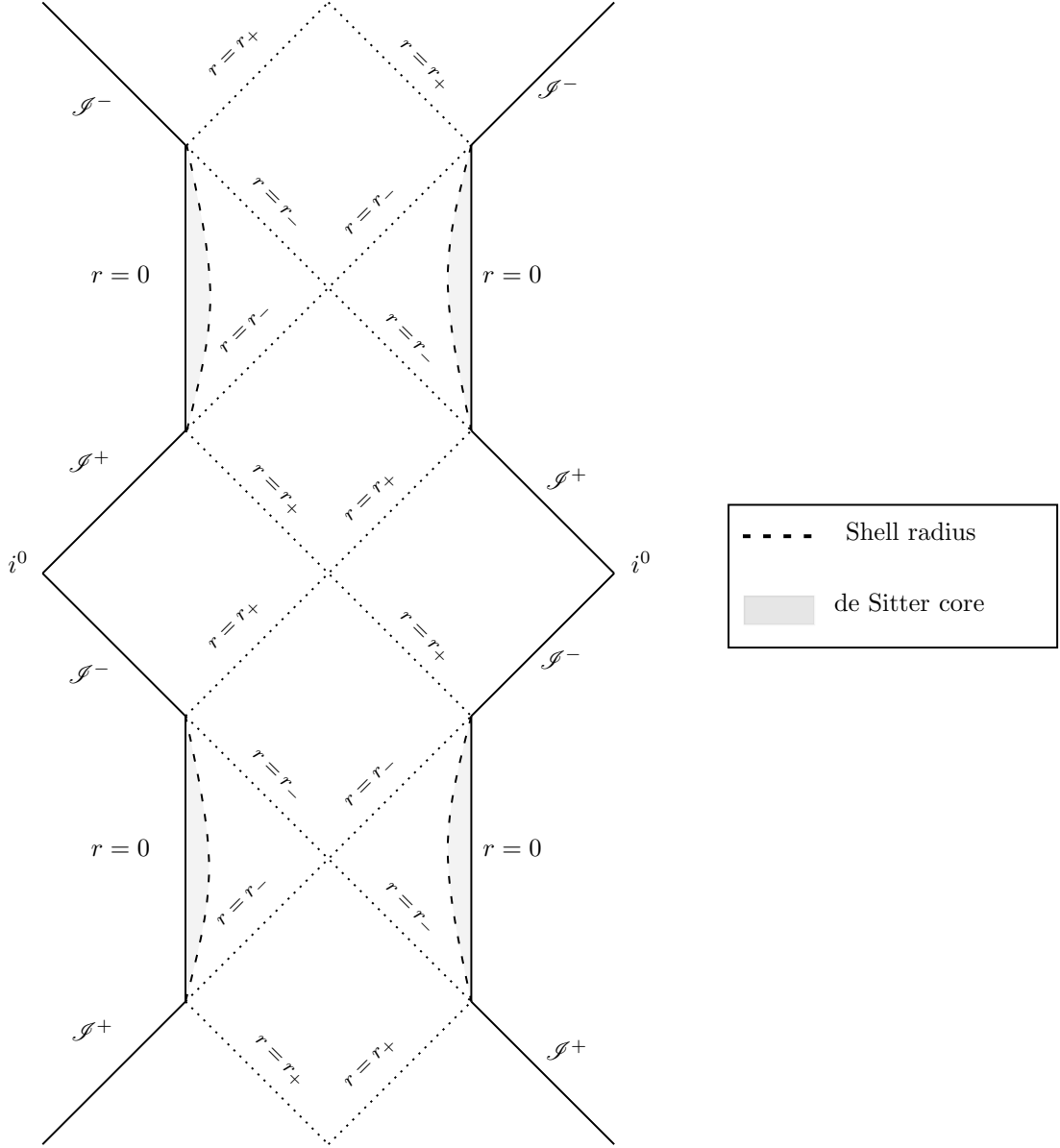


\section{Concluding Remarks}\label{sec: Concluding Remarks}
In this paper, it is shown that Kiselev BH has at most two horizons which both are smaller (larger) than Schwarzschild horizon if $\omega>0$ and $\tilde{c}<\tilde{c}_{ext}$ ($\omega<0$ and $\tilde{c}>\tilde{c}_{ext}$). To study thin shell gravitational collapse in Kiselev geometry, we have  considered first, a null thin shell with Minkowski core and then a timelike thin shell with de Sitter core.

In the former case, as expected, one can calculate the surface energy-momentum tensor by determining the Einstein tensor using ingoing Eddington-Finkelstein coordinates without appealing to the Barrabes-Israel junction conditions.

In the latter case, we have shown that the equation of state parameter must be positive because of the requirement that the shell is timelike and the energy density is positive. Invoking Barrabes-Israel junction conditions we found out that stationary BH solutions can be found from equations (\ref{eq: set1 a}) and (\ref{eq: set1 b}) where we have concluded that the normalized shell mass should satisfy $M<1/2$.  Stability of these solutions have been examined numerically and it is shown that for $\omega<1/3$, there exists stable BH solutions with negative shell mass which is unbounded from below. This is in contrast with the case of a charged regular BH constructed in \cite{Uchikata:2012zs} where solutions with negative shell mass are unstable. Moreover, in our solutions, the free parameters of Kiselev BH, $m$ and $\abs{c}$ are unbounded from above. However, solutions with negative shell mass are not physically acceptable. Considering the BH solutions with positive shell mass,  we have found the allowed ranges of parameters of the shell and BH that yield stable stationary BH configurations and the results is presented in figure \ref{fig: mwr}. For example, setting $\omega=0.1$, the valid ranges of parameters are given by $0\lessapprox M\lessapprox0.42$, $1.9\lessapprox m\lessapprox5.4$ and $-9.80\lessapprox c\lessapprox-2.81$ while the normalized shell radius lies on $0.84\lessapprox R\lessapprox0.99$. Moreover, this can be done even analytically for the particular choice $\omega=1/3$. In this case, our results are approximately the same as those obtained by \cite{Uchikata:2012zs}.

\vglue1cm
{\bf Acknowledgments:}

The authors would like to thank the Iran National Science Foundation (INSF) for supporting this research under grant number 99000365. F. Shojai is grateful to the University of Tehran for supporting this work under a grant provided by the university research council. We are very grateful to the 
anonymous Referees for the evaluation of our paper and for the constructive critics.

\end{document}